# Obliquity Dependence of the Formation of the Martian Polar Layered Deposits

Jeremy A. Emmett[a*]

James R. Murphy[a]

Melinda A. Kahre[b]

[a] Department of Astronomy, New Mexico State University, Box 30001, MSC 4500, Las Cruces, NM 88003-80001, USA

[b] Space Science and Astrobiology Division, NASA/Ames Research Center, Moffett Field, CA 94035, USA

* Corresponding author. Email address: jeem9157@nmsu.edu

Current postal address: Department of Astronomy, New Mexico State University, Box 30001, MSC 4500, Las Cruces, NM 88003-80001, USA.

**Abstract**

Mars' polar layered deposits (PLD) are comprised of layers of varying dust-to-water ice volume mixing ratios (VMR) that are thought to record astronomically-forced climatic variation over Mars' recent orbital history. Retracing the formation history of these layers by quantifying the sensitivity of deposition rates of polar material to astronomical forcing is critical for the interpretation of this record. Using a Mars global climate model (GCM), we investigate the sensitivity of annual polar water ice and dust surface deposition to a variety of obliquity and surface water ice source configurations at zero eccentricity that provide a reasonable characterization of the evolution of the PLD during recent low-eccentricity epochs. The GCM employs a fully interactive dust lifting/transport scheme and accounts for the coupling effects of the physics of dust and water on the transport and deposition of water ice and dust. Under a range of tested obliquities (15° - 35°), the predicted net annual accumulation rates range from -1 mm/yr to +14 mm/yr for water ice and from 0.003 - 0.3 mm/yr for dust. GCM-derived accumulation rates are ingested into an integration model that simulates polar accumulation of water ice and dust over five consecutive obliquity cycles (~700 thousand years) during a low eccentricity epoch. A subset of integration model simulations predict combined north polar water and dust accumulation rates that correspond to the observationally-inferred average growth rate of the north PLD (0.5 mm/yr). These integration model simulation results are characterized by net water transfer from the south to the north polar region. In the north, a ~230 m-thick deposit is accumulated over ~700 thousand years. Three types of layers are produced per obliquity cycle: a ~30 m-thick dust-rich (20 - 30% dust VMR) layer that forms at high, a ~0.5 m-thick dust lag deposit (pure dust) that forms at low obliquity, and two ~10 m-thick dust-poor (~3%) layers that separate the dust rich layers and form when obliquity is increasing or decreasing. The ~30 m-thick dust-rich layer is reminiscent of a ~30 m scale length feature derived from analysis of visible imagery of north PLD vertical structure, while the ~0.5 m-thick dust lag is only a factor of ~2 smaller than observed "thin layers". Overall, this investigation provides further evidence for an obliquity forcing in the PLD climate record, and demonstrates the importance of ice-on-dust nucleation in polar depositional processes.

## 1. Introduction: Mars Polar Layered Deposits

Mars' north and south poles are covered in deposits of ice and dust referred to as Polar Layered Deposits (PLDs). Shallow-sloped troughs that cut into the upper ~1 km of these deposits reveal a series of alternating dark and bright lines running laterally along the strike of their walls, interpreted as layers of icy material with varying dust-to-water ice VMRs and/or other properties that could give rise to varying albedo with depth (Laskar et al., 2002). High resolution imagery obtained by the Mars Orbital Camera (MOC) on board Mars Global Surveyor (MGS) (Malin and Edgett, 2001; Kolb and Tanaka, 2001) and the High Resolution Imaging Science Experiment (HiRISE) onboard Mars Reconnaissance Orbiter (MRO) (McEwen et al., 2007) reveal that individual layers range between decimeters to tens of meters in thickness and appear to vary in texture and erosional resistance (Milkovich and Head 2005; Fishbaugh et al. 2006; Fishbaugh et al., 2010a,b; Limaye et al. 2012; Becerra et al., 2017, 2019). However, the mantling of decimeter-scale layers in trough exposures by sublimation dust lag deposits may obscure thinner layer structure (Herkenhoff et al., 2007). Radar cross-sections obtained by MRO's SHallow RADar sounder (SHARAD; Seu et al. 2007) and Mars Express' Mars Advanced Radar for Subsurface and Ionosphere Sounding (MARSIS; Picardi et al., 2005) suggest that layering sequences within the deposits extend to ~2 km in depth in the north and up to over (Phillips et al. 2008) 3 km in depth in the south (Plaut et al. 2008), and are laterally continuous on scales of tens to hundreds of kilometers, with some local unconformities (Milkovich and Plaut, 2008, Putzig et al. 2009). Radar reflectance data from SHARAD suggest a bulk average dielectric constant of 3.1 for the deposits, implying a primarily water ice composition, with bulk volume dust-to-water ice concentrations of ~5% in the North PLD (NPLD) (Phillips et al., 2008; Grima et al., 2009) and ~10% for the South PLD (SPLD) (Plaut et al., 2007). The PLD extend equatorward to 80° in the north and 70° in the south (P Thomas et al., 1992). Age estimates suggest the PLD to be relatively young, ~5 Million years (5 Myrs) in the north (Montmessin, 2006; Levrard et al. 2007; Hvidberg et al., 2012) and perhaps 10 - 100 Myrs in the south (Plaut et al., 1988; Herkenhoff and Plaut, 2000, Koutnik et al. 2002).

The presence of the PLD at latitudes currently conducive to seasonal ice and dust accumulation, and their extensive, layered structure, suggest that they may have formed via the net deposition of atmospherically-transported water and dust (e.g. Cutts et al., 1973). Previous modelling indicates that atmospheric water and dust abundances, atmospheric transport, and the stability of surface water ice in the polar regions are sensitive to orbit and obliquity cycle-induced insolation variations. (Mischna et al., 2003; Newman et al., 2005; Forget et al., 2006; Levrard et al., 2007; Madeleine et al., 2009, 2014). It has been proposed that insolation variations could drive the formation of the PLD by periodically altering the net annual flux of those aerosols into and out of the polar regions (e.g. Laskar et al., 2002). Observational evidence has since supported such an interpretation. Investigations have identified quasi-periodic wavelengths in the vertical spacing between layers at various spatial scales, and individually correlated these with single-period oscillations in Mars' calculated obliquity and precession histories (Laskar et al., 2002; Milkovich and Head, 2005; Fishbaugh et al., 2010a; Limaye et al., 2012). Most recently, time-series analysis of exposed layer sequences in the PLD has revealed a close correlation between the ratio of two dominant stratigraphic wavelengths and that of two dominant periods of polar insolation variability driven by from astronomical forcing (Becerra et al., 2017, 2019).

Yearly-averaged insolation at a given latitude is primarily modulated by the planet's obliquity (Laskar et al., 2002), which has oscillated between 15° and 45° over 120,000 yr. cycles during the past ten million years (Fig. 1). Additionally, orbital eccentricity has oscillated between 0 and 0.12 over ~100 thousand year (kyr) cycles (Fig. 1), and precession of the longitude of perihelion cycles on 51 kyr timescales.

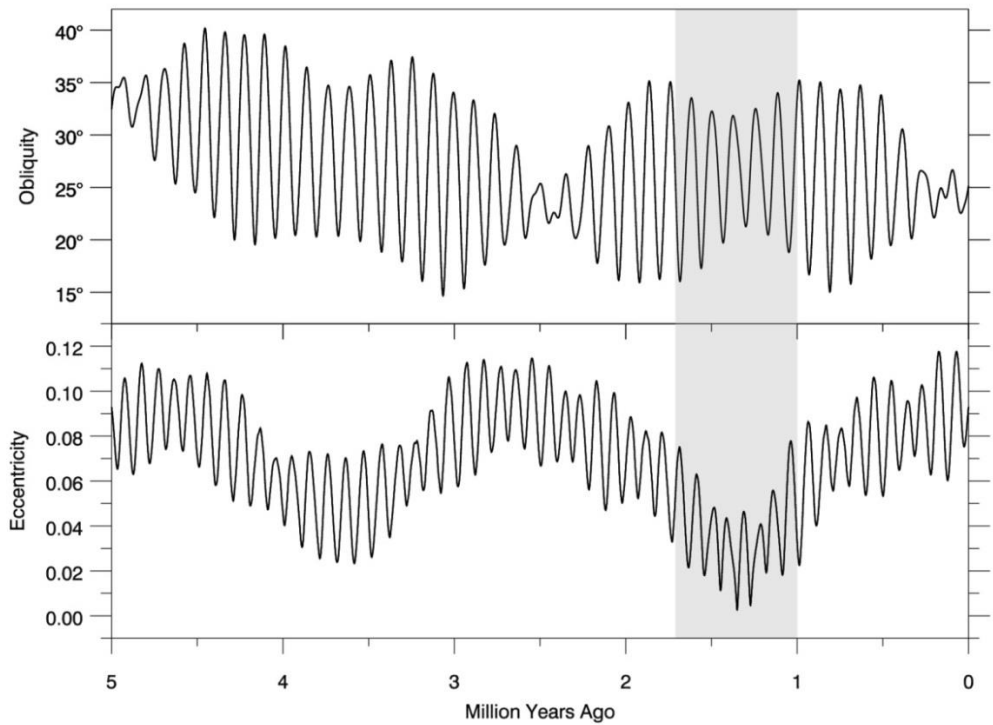

**Fig. 1.** Mars' obliquity and eccentricity histories from 5 million years ago to the present-day. The Mars atmosphere model simulations discussed in sections 3 and the integration model simulations discussed in section 4 relate to the low eccentricity period from ~1.7 to ~1 million years ago (gray strip). Data are derived from Laskar et al. (2004) – the north polar surface insolation history resulting from these orbital and axial oscillations is displayed in Figure 1 of that work.

Investigations based on Martian climate models have explored the effects of varying orbital and axial parameters on the global atmospheric redistribution of dust and water surface reservoirs. Newman et al. (2005) utilized the Oxford-LMD (Laboratoire de Météorologie Dynamique) Global Climate Model (GCM) to investigate the influence of obliquity variations on Mars' dust cycle. Annual dust accumulation was predicted to occur in the polar regions at all tested obliquities. These simulations utilized interactive surface wind stress and dust devil lifting parameterizations. Permanent $CO_2$ polar ice caps that develop under low obliquity conditions reduce the near surface gas density, reducing surface wind stress dust lifting and corresponding polar accumulation rates of dust. Increased near-surface atmospheric gas density and wind speeds under higher obliquity conditions result in a dustier atmosphere that drives additional lifting in a positive feedback cycle induced by the radiative effects of suspended dust in the atmosphere.

The effect of obliquity and orbital variations on the annual redistribution of surface water ice was investigated by Mischna et al. (2003), Forget et al. (2006), Levrard et al. (2007), and Madeleine et al. (2009, 2014). Levrard et al. (2007) used the LMD GCM to investigate how the evolution of Mars' orbit and axial parameters over the past 10 Myr (Laskar et al. 2002) may have driven the exchange of surface water ice between polar and non-polar reservoirs. In these simulations, water ice was annually transferred from the north polar region to the low and midlatitudes for annually-averaged north polar surface insolation values > 298 $W/m^2$, corresponding to obliquities of 35° - 45° depending on orbit/eccentricity configuration.

Madeleine et al. (2009) predicted the accumulation of a northern midlatitude surface water ice reservoir from a tropical source at moderate obliquities (25° - 35°). Similarly, Madeleine (2014) predicted that a northern midlatitude surface water ice reservoir would accumulate from a north polar cap source under these obliquity conditions. The net annual accumulation rate of this midlatitude ice is maximized in high eccentricity epochs (e = 0.1) and minimized in low eccentricity epochs (e = 0), implying interglacial periods may occur during low eccentricity epochs on Mars.

High obliquity (45°) models have predicted a migration of water ice from the polar regions to tropical locations (Forget et al., 2010), with accumulation preferred on the western flanks of the Tharsis volcanos when water ice is sourced from a northern cap, and east of Hellas Basin when water is sourced from a southern cap. Predicted regions of surface water ice accumulation outside of the polar regions are consistent with the presently-observed locations of glacial-like landforms: midlatitude lobate debris aprons (LDA) (Squyres et al., 1992; Pierce and Crown, 2003; Chuang and Crown, 2005), midlatitude lineated valley fill (LVF)/plateau glaciation (Head et al., 2006a,b, 2010; Fastook et al., 2011, 2014), mid-to-high-latitude concentric crater fill (CCF) (Kreslavsky and Head, 2006; Levy et al., 2010; Fastook and Head, 2014), midlatitude ice highstands (Dickson et al., 2008, 2010), low to midlatitude phantom lobate debris (Hauber et al., 2008), and tropical mountain glaciers (Forget et al., 2017).

The GCM-based investigations described above demonstrate that the martian dust and water cycles individually exhibit substantial sensitivity to Mars' orbit and obliquity state, but the effect of their physical interactions on polar deposition or removal has yet to be thoroughly investigated with GCMs. Dust lag deposits may impede sublimation of dust-covered ice, preserving and/or enhancing the growth rate of water and $CO_2$ ice deposits (Levrard et al., 2007; Bramson et al., 2019). Surface water ice cover may impede dust lifting, particularly under high obliquity conditions characterized by latitudinally-extensive seasonal water ice polar caps (Mischna et al., 2003). Suspended dust particles nucleate to form clouds (Haberle et al, 2019; Nelli et al., 2009; Montmessin et al. 2002), likely inhibiting cross-equatorial dust transport (Clancy et al., 1996), and dust-nucleated cloud formation enables the co-deposition of dust with water ice particles. Spatial and temporal variability in the distribution of atmospheric dust is well characterized for Mars' present climate, but probably differs considerably and non-linearly under different orbital and obliquity forcing (e.g. Newman et al. 2005). Previous paleoclimate water cycle investigations have considered different atmospheric dust opacity regimes, but for simplicity, have generally used time invariant dust opacities (Madeleine et al., 2009; Forget et al., 2010) rather than a self-consistently evolving dust opacity/distribution. The investigation of the dust cycle by Newman et al. (2005) utilized such an evolving atmospheric dust distribution but excluded the presence of water.

Analytical and GCM-based models of PLD formation have demonstrated the importance of contemporaneously quantifying dust and water ice deposition in order to reproduce vertical layered water structure. Hvidberg et al. (2012) modelled variations in the net annual polar deposition rates of both water ice and dust at the north pole over the past 10 Mya using a simplified deposition/removal parameterization based on Mars' calculated insolation record. Resultant stratigraphies of layer thickness and dust-to-water ice VMR vs. depth and time were compared to observed stratigraphic columns in the PLD. This model demonstrated the ability of dust-rich icy layers in the PLD to form via two possible mechanisms: 1) high rates of dust deposition relative to water ice deposition when both are being simultaneously deposited and 2) formation of dust lag deposits during periods of net ice ablation or dust deposition alone. Though such models self-consistently determine the dust concentration of deposited material and thus eliminate the need to assume an initial PLD dust concentration (Levrard et al., 2007), they parameterize rather than explicitly account for the effects of 3D transport and dynamics in the determination of net annual polar deposition/removal rates.

As a logical next step, the present work investigates PLD formation using a GCM to simulate coupled water and dust cycles (via a water ice cloud microphysics scheme (Haberle et al., 2019) under varying martian obliquity and zero eccentricity. We focus on a time period 1.7 - 1 million years ago (Ma) - when eccentricity was small (Fig. 1) - to create a manageable parameter space to investigate. The GCM allows the water and dust distributions to interactively evolve. This accounts for sublimation from prescribed and model-created surface ice reservoirs, water ice particle nucleation on suspended dust and possible resultant snowfall, lifting from/deposition of dust on the surface, inhibition of dust lifting by surface ice, and frosting of the surface.

## 2. Mars Atmosphere Model

In this investigation, we utilize the NASA-Ames Legacy Mars Global Climate Model (MGCM) (Haberle et al., 2019) to investigate the sensitivity of polar surface water ice and dust accumulation rates to changes in Mars' obliquity and surface water ice source locations. The MGCM numerically solves the primitive equations of meteorology in three dimensions and implements a radiatively-active dust cycle coupled to the water cycle (Kahre et al., 2015). The MGCM microphysics package simulates ice particle nucleation upon suspended dust, ice particle growth or shrinkage, and precipitation (Montmessin et al. 2002; Nelli et al., 2009; Haberle et al., 2019). A two stream radiation code allows for gas and dust opacity treatment in both the visible (quadrature) and the IR (hemispheric mean), deriving gas opacities from $CO_2$ and $H_2O$ k-coefficient data and dust opacity from Wolff dust optical properties (Mischna et al., 2012). The radiative effects of water ice clouds and $CO_2$ clouds are not accounted for in this investigation.

Cloud radiative effects would be an additional factor requiring specific interpretation beyond our obliquity-focused analysis, which already takes a significant step forward in its assessment of the impact of the water and dust microphysical coupling. The GCM for this investigation was configured with a 5° x 6° latitude-longitude grid and 24 vertical normalized pressure levels. This enables analysis of geographic trends in the magnitudes of surface accumulation rates between 85° South to 85° North and analysis of atmospheric aerosol fluxes from the near surface to ~80 km altitude. The "polar regions" analyzed in this work refer to surface locations poleward of the 77.5° North and South latitude circles, which roughly delineate the edges of the present-day PLD. The values for atmospheric composition and solar luminosity are those of the present-day. These are not expected to have varied significantly over the past 10 Myrs.

The GCM accounts for water vapor, water ice, non-nucleated ("free") dust, and nucleated ("core") dust tracers, with water ice and dust particle sizes conforming to assumed log-normal size distributions with prescribed standard deviations for each species (Haberle et al., 2019). The model utilizes a 'moments' scheme for tracer transport (Haberle et al., 2019) that allows the instantaneous particle size distribution of each tracer to be fully described by two transported quantities: the mass and number of particles per unit volume.

## 2.1 Water and CO$_2$ Ice Sublimation and Deposition

Direct condensation/sublimation of H$_2$O and CO$_2$ onto/from the surface occurs at rates determined by local vapor pressure and temperature conditions. H$_2$O can additionally be precipitated onto the surface as snow, with a rate determined by particle-size dependent settling velocities. Condensation and sublimation rates vary according to a bulk transfer equation derived from Brutsaert et al. (1982) and described by equations (1) & (2) of Haberle et al. (2019), with the rates dependent upon the local saturation mixing ratio (which is a function of temperature), on the near-surface wind speed / friction velocity), and on the local surface roughness.

At all locations that are not initial prescribed source regions of surface water ice, an albedo feedback scheme is employed in which the surface albedo is set to 0.4 (bright water ice) if the thickness of deposited water ice exceeds 5 μm. Although the ice albedo is expected to be inversely proportional to its dust content, and darkening due to dust may enhance sublimation rates of ice, we have neglected this effect due to difficulties in 1) dynamically tracking the time evolution of the surface ice deposits' dust fractions and 2) translating dust content into albedo. All source regions are permanently prescribed a constant albedo and thermal inertia of 0.35 and 800 J m$^{-2}$ s$^{-1/2}$ K$^{-1}$, respectively, similar to the present spatially-averaged surface properties of the north residual ice cap (Navarro et al., 2014; Madeleine et al., 2014).

## 2.2  Dust Lifting and Deposition

Dust is initially-lifted from the surface with a prescribed log-normal particle size distribution providing an average particle radius of 1.5 μm, consistent with observation-derived atmospheric dust particle sizes (Wolff and Clancy, 2003). Atmospheric water vapor condenses on suspended dust particles, forming a nucleated dust core tracer (termed "core dust") that is thus equal in number density to water ice particles. Dust cores are contemporaneously deposited when water ice precipitates, but the model prevents deposited core dust from subsequently being lifted from the surface (i.e., the model assumes permanent "entrapment" of this core dust on the surface) . Free dust may be lifted from any surface location. However, for the locations where prescribed water ice is sourced from the surface, $CO_2$ ice is present on the surface in any quantity, or deposited water ice is present on the surface in excess of 10 μm, a reduction in surface roughness by a factor of 100 is applied that significantly reduces surface wind stress lifting.

Free dust is deposited on the surface via sedimentation (determined by particle-size dependent settling velocities of free dust), and is removed from the surface via surface wind stress and dust devil lifting. The MGCM accounts for two fully-interactive dust lifting schemes: (1) a surface wind stress dust lifting scheme (Kahre et al., 2006; Haberle et al., 2019) dependent on a surface wind stress threshold (22.5 mN m$^{-2}$) that must be exceeded to initiate lifting and a lifting efficiency parameter (0.047), and (2) a dust devil lifting scheme (Rennó et al., 1998; Newman et al., 2002; Haberle et al., 2019) dependent on the magnitude of the heat exchange between the surface and atmosphere, the depth of the planetary boundary layer, and a separate lifting efficiency parameter (1 x 10$^{-10}$). The threshold and two parameters related to dust lifting were tuned to provide reasonable agreement between the observed and coupled-modelled dust and water cycles when the model is configured to simulate current Mars conditions. All simulations described in this work utilize the same lifting parameterization for consistency, regardless of obliquity or initial surface water ice distribution.

Water and dust mass are not conserved within the MGCM simulations. The maximum impact of this conservation issue does not change the qualitative results of this investigation, and is addressed in sections 3 and 4.

## 2.1 Simulations

### 2.1.1 Orbit and Obliquity Parameters:

We simulate orbit/obliquity/climate conditions of interest for this investigation by specifying Mars' orbit and obliquity parameters. Because polar insolation is most strongly determined by obliquity, this investigation focuses solely on the effect of varying obliquity. The orbital eccentricity is set to zero to circularize the orbit in all simulations, removing the hemispheric bias that would result from alignment of a particular season with perihelion passage. Topographic variations are expected to provide a hemisphere-dependent forcing (Richardson and Wilson 2002). In this obliquity-focused investigation, specified obliquity values span the predicted range during the past 10 Mya (15° - 35°) (Laskar et al., 2004). MGCM simulations were run for 20 simulated Mars years to stabilize their $CO_2$, water, and dust cycles. While eccentricity is expected to have an effect on longer timescales and perhaps explain differences between the north and south PLD, investigation of the impact of the eccentricity parameter would greatly broaden the investigation parameter space and is beyond the scope of this work.

### 2.1.2 Initial Surface Properties and Surface Water Ice Distribution:

Previous investigations of martian surface water ice stability under different orbital configurations predict permanent polar water ice deposits under low obliquity conditions, and polar ice loss and accumulation at midlatitudes during moderate obliquity (e.g. Levrard 2007). The MGCM utilizes present-day, global maps of surface albedo (Putzig and Mellon, 2007), thermal inertia (Putzig and Mellon, 2007), and topography (Smith et al, 1999). To avoid imposing modern-day surface properties in the polar regions when simulating past orbital epochs (and possibly biasing simulated annual water ice and dust accumulation), the polar surfaces (77.5° to 87.5° N/S) are assigned initial spatially constant values in surface albedo, thermal inertia, and topography equal to their nominal 67.5° to 72.5° N/S zonal averages (the present-day polar ice caps are effectively removed by making the polar surfaces resemble surrounding dry terrain). Although the basal topographies of the PLD have been mapped via radar sounding (Holt et al., 2010; Selvans et al., 2010; Brothers et al., 2015; Nerozzi et al., 2018; Putzig et al., 2018), we favor the prescription of spatially-constant polar surface properties for the following reasons: a small thermodynamic effect is expected to result from the addition of up to ~1 km of topographic relief, the MGCM's 5° x 6° latitude/longitude grid poorly resolves variations in polar surface topography, and uncertainties remain regarding the surface albedo and thermal inertia of the basal deposits.

The locations on the surface where water ice is initially emplaced (i.e., where water is sourced) are specified to simulate hypothetical, idealized distributions of surface water ice that are characteristic of conditions within the past five million years (Head et al., 2003; Mischna et al., 2003; Levrard et al., 2007; Madeleine et al., 2009, 2014; Forget et al., 2017). Previous investigations have predicted the formation of permanent polar water ice caps at low obliquities (~15°) and the formation of middle latitude surface water ice deposits at moderately-high obliquities (~35°) (e.g. Mischna et al., 2003; Levrard et al., 2007; Madeleine et al., 2009). This investigation focuses on these two predicted surface water ice distributions. For sources of water at the poles, the effect of emplacing water ice caps simultaneously in both polar regions is investigated (i.e., poleward of 77.5° N/S). For midlatitude sources, the effect of simultaneously emplacing planet encircling strips of water ice in the north and in the south midlatitudes is investigated (surface water ice is emplaced within 5°-latitude-wide bands spanning 37.5° - 42.5° N/S). These ranges are chosen because they encompass a high concentration of several types of observed water-ice related features: CCF, LVF, LDA, and Glacial Highstands (Forget et al., 2017). Furthermore, the sub-kilometer topography of these regions exhibits a relatively high degree of dissection and a low degree of topographic concavity compared to other latitudes, implying the emplacement and subsequent removal of a meters-thick deposit of water ice within the past two million years that is thought to be associated with the present latitude-dependent mantle (LDM) deposit (Head et al., 2003). A planet-encircling band of water ice is emplaced ( rather than a longitudinally-varying distribution of surface water ice) due to uncertainty in the location and extent of the midlatitude water ice deposits that would have produced the observed features. An albedo feedback scheme is employed that dynamically sets the surface albedo to 0.4 (clean ice) where surface water ice exceeds 5 μm in thickness or where $CO_2$ ice is present on the surface in any quantity. While recent observations suggest likely higher $CO_2$ ice albedos of 0.6 – 0.9 in the south residual $CO_2$ cap (James and Wolff, 2018), the measured Viking pressure/CO2 cycle is reasonably reproduced by the MGCM with the described feedback scheme (Haberle et al., 2019). In a 'baseline' present-day simulation, the GCM's default surface properties and imposed north polar residual water ice cap source are employed (the polar surfaces aren't

'smoothed') with the albedo feedback scheme only applied to non-residual cap locations that acquire water ice. Under present-day conditions, the MGCM does not consider the presence of the $CO_2$ south residual cap, which would act as a permanent sink of water via cold-trapping (Mischna et al., 2003; Haberle et al., 2019).

## 2.2 Quantifying Net Annual Polar Water Ice and Dust Deposition Rates

In each simulation, annual rates of surface water ice and dust deposition or removal within the polar regions are quantified by calculating the average yearly change in surface water ice and dust mass between simulated years 15 and 20, representing a long-term trend that accounts for inter-annual variation. In years 15 and 20, at every location on the surface, the mass density ($kg/m^2$) of surface water ice and dust are measured at the date when the surface water ice abundance is minimized in local summer. Though this precise date during local summer may vary annually and geographically, this method is preferred over quantifying mass densities at a fixed date each year, because it mitigates the possibility that the calculated annual changes are influenced by inter-annual variations in small-magnitude seasonal frost. The annual change in mass is converted to a change in thickness, dividing it by assumed 'packing densities' (2000 $kg/cm^3$ for porous dust and 900 $g/cm^3$ for water ice). All polar accumulation or loss rates reported in this work are expressed as "per Earth year".

Assuming that material deposited over annual timescales is uniformly intermixed, the dust-to-water ice VMR of the deposited material is calculated as the ratio of the dust thickness to the dust + water ice thickness. A mixing ratio of zero describes pure water ice (no intermixed dust), a mixing ratio of 1 describes pure dust (no intermixed water, also referred to as a "dust lag" deposit in this work), and a mixing ratio between 0 and 1 describes a mixture of dust and water ice. Values closer to 1 are for more "dust-rich" deposits.

## 2.3 Seasonal Considerations

Analyzing seasonal changes in the vertical structure of simulated polar surface deposits provides insight into processes that determine the deposit's multi-year average growth rate and dust-to-water ice VMR. This investigation focuses on the deposition and removal of surface water ice and dust in the polar regions. Although seasonal deposition and sublimation of $CO_2$ does occur and would in reality influence the seasonal evolution in the deposit vertical structure, it generally does not impact the annual change in thickness of the water ice and dust. All simulations apart from the 15° obliquity polar cap source simulation exhibit a total summertime depletion of seasonally-deposited $CO_2$ ice, resulting in no annual change in surface $CO_2$ ice thickness and making dust and/or water ice the only constituents of annually-accumulated PLD layers.

## 3. Simulation Results

We ran 55 GCM simulations to investigate how the variation of obliquity and location of water sources affects the observed vertical structure of the PLDs. Resultant annual polar accumulation rates and dust-to-water ice VMRs for each of these scenarios are summarized in Table 1 of section 4. We present six of these simulations, representing prescribed initial polar cap or midlatitude surface water ice sources under 15°, 25°, and 35° obliquity and zero eccentricity conditions. Additionally, a 'baseline' simulation is analyzed that was calibrated to approximate the present climate (observed dust and water cycles), utilizing the present orbit (eccentricity 0.093; longitude of perihelion 251°) and obliquity parameter (25.19°), and an initial surface water ice distribution consisting of a northern permanent water ice cap that is defined by the morphology of the present-day north residual cap (Haberle et al., 2019).

## 3.1 Present-Day Conditions

We first present results from a simulation approximating Mars' present dust, water, and $CO_2$ cycles, i.e., when the MGCM is configured with Mars' present orbital parameters and a present north residual water ice cap as an initial water source. We utilize observationally-derived global maps of surface albedo, thermal inertia, and topography (derived from Mars Observer Laser Altimeter topography smoothed to a 5° latitude x 6° longitude resolution (Haberle et al., 2019) to accurately represent Mars' polar surfaces in the present epoch.

The seasonal evolution of the simulated zonally-averaged water vapor column abundance (Fig. 2a) well reproduces MGS/TES Mars year 24 - 27 observations. Sublimation of polar water ice produces the north polar region's zonally averaged water vapor column abundance maximum ~50 pr. μm around $L_s$ ~105°, and the south's at ~20 pr. μm around $L_s$ ~255°. The simulated southern maximum occurs 30 degrees of $L_s$ earlier ($L_s$ 255°) than observed ($L_s$ 285°) despite producing an observationally-consistent water vapor column maximum.

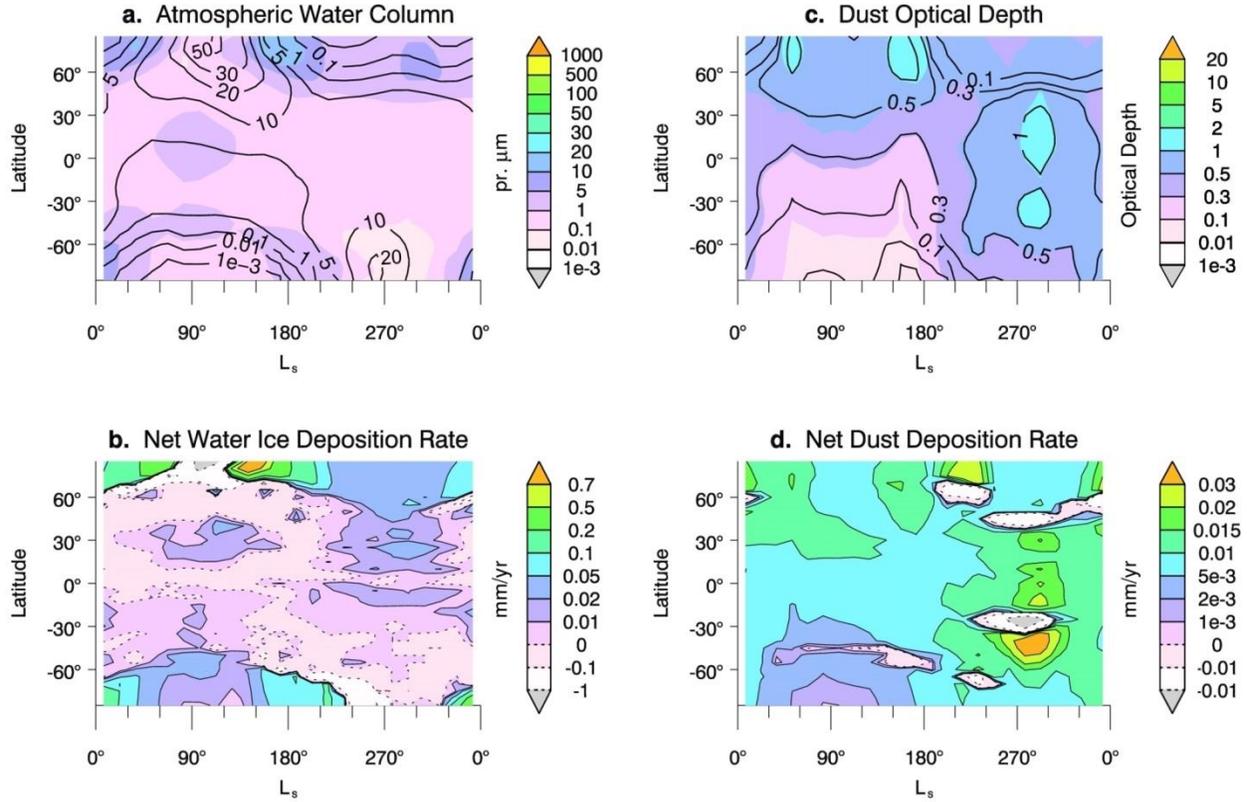

**Fig. 2. Present-Day Simulation Water and Dust Cycles. a.** Water ice column (colored contours, pr. μm) and water vapor column (over-plotted contours, pr. μm) as a function of latitude and time of year. **b.** Net water ice deposition rate (mm/yr) as a function of latitude and time of year, with solid/dashed lines enclosing positive/negative rates. Net deposition rates are expressed as "per Earth year". **c.** Total visible dust optical depth (colored contours) and free dust visible optical depth (over-plotted contours) as a function of latitude and time of year. **d.** Net dust deposition rate (mm/yr) as a function of latitude and time of year, with solid/dashed lines enclosing positive/negative rates.

There is no net transfer of water from the prescribed northern cap source to the south polar region. There is an annual redistribution of water ice from the periphery of the prescribed northern cap to the cap interior, consistent with the results of Haberle et al. (2019) and Navarro et al. (2014).

The interior of the northern source region (poleward of 77.5° N) accumulates water ice at 0.06 mm/yr at the expense of water ice lost from the source periphery equatorward of 77.5° N. In both polar regions, net water accumulation rates (Fig. 2b) are maximized in early spring and late summer, are annually minimized from early autumn through late winter, and counteracted by sublimation-driven net removal of water ice around the time of summer solstice. This is dynamically consistent with Richardson and Wilson (2002) in terms of seasonal variations in high-latitude atmospheric water content and the expansion and contraction of turbulent polar vortex walls. The south polar region does experience some seasonal water ice accumulation, but that seasonally accumulated water ice completely sublimes at the remaining times of the year. In the north, sublimation at the time of local summer solstice removes some water ice accumulated in early spring and late summer. Yearly-integrated annual water ice deposition is almost entirely driven by snowfall in both polar regions (> 90%) rather than by frost.

Seasonal variations in simulated dust optical depth are generally consistent with observations obtained by MGS/TES and ODY/THEMIS for MY 24 - 32 (Smith et al. 2008; Kahre et al., 2017). Values (Fig. 2c) vary between ~0.1 to ~0.5 during northern spring and summer, increase to ~1 at northern fall equinox in response to regional dust storms at high and low midlatitudes, and remain enhanced from southern summer through fall. As noted in previous fully-interactive dust cycle investigations (see Kahre et al., 2017), the simulated dust activity produces a single peak in globally-averaged dust opacity at $L_s$ ~285°, which is inconsistent with the two peaks in globally average dust opacity that are observed to occur at $L_s$ ~240° and ~340°.

The north and south polar regions annually accumulate dust at rates of 0.01 mm/yr and 0.005 mm/yr, respectively, with deposition of dust cores dominating the dust component of the annual deposit (~85% of deposited dust in the north and ~65% in the south). Maximum seasonal rates of polar dust deposition occur in early-to-mid local spring and autumn, most prominently in the north (Fig. 2d). It is notable that the seasons of maximum polar dust accumulation do not precisely coincide with the seasons of maximum water ice accumulation (Fig. 2d) despite dust being delivered to the polar surfaces as snow particle nuclei in all seasons except around local summer solstice. This result arises from seasonal variation in the dust-to-water ice VMR of deposited snow particles, which exhibits a strong correlation with rates of seasonal dust deposition and is similarly maximized in early-to-mid local spring and autumn (ranging from ~1% to 15%). Dust lifting poleward of 77.5° N/S is permitted to occur, but such lifting would be infrequent and is generally insignificant compared to simultaneous dust deposition, resulting in year-round net positive dust deposition in both polar regions (Fig. 2d).

The interior of the northern water ice source region (poleward of 77.5° N) accumulates an annual ~0.7 mm "layer" of water ice and dust with a dust-to-water ice VMR of ~14%. The south polar surface, which accumulates dust but does not retain water ice, accumulates a 0.005 mm-thick dust covering (Fig. 3a). The seasonal growth and evolution of the polar region's water ice and dust deposits is primarily driven by the seasonal evolution of water ice deposition, consistent with the results of Hvidberg et al. (2012), Brown et al. (2016), and Becerra et al. (2007). Summer solstice sublimation of surface dust-containing water ice and retention of previously-deposited core dust, and simultaneous air fall of free dust forms dust lag deposits (dark orange layers in Fig. 3.7). This seasonal loss of water ice and the retention/accumulation of dust results in year-end bulk polar VMR's in the north (14%) and south (100%) that are greater than the polar regions' respective average snow particle VMRs of ~2% and ~5%. In all simulations conducted in this investigation, summertime sublimation of water ice results in a larger annual deposit bulk VMR compared to the average VMR of locally precipitated snow particles. This effect is especially important among the present-day and paleo-climate polar cap scenarios as they exhibit the most substantial summertime sublimation of water ice (sufficient to drive net removal of water and form a seasonal dust lag deposit) and simultaneous free dust deposition in the same season.

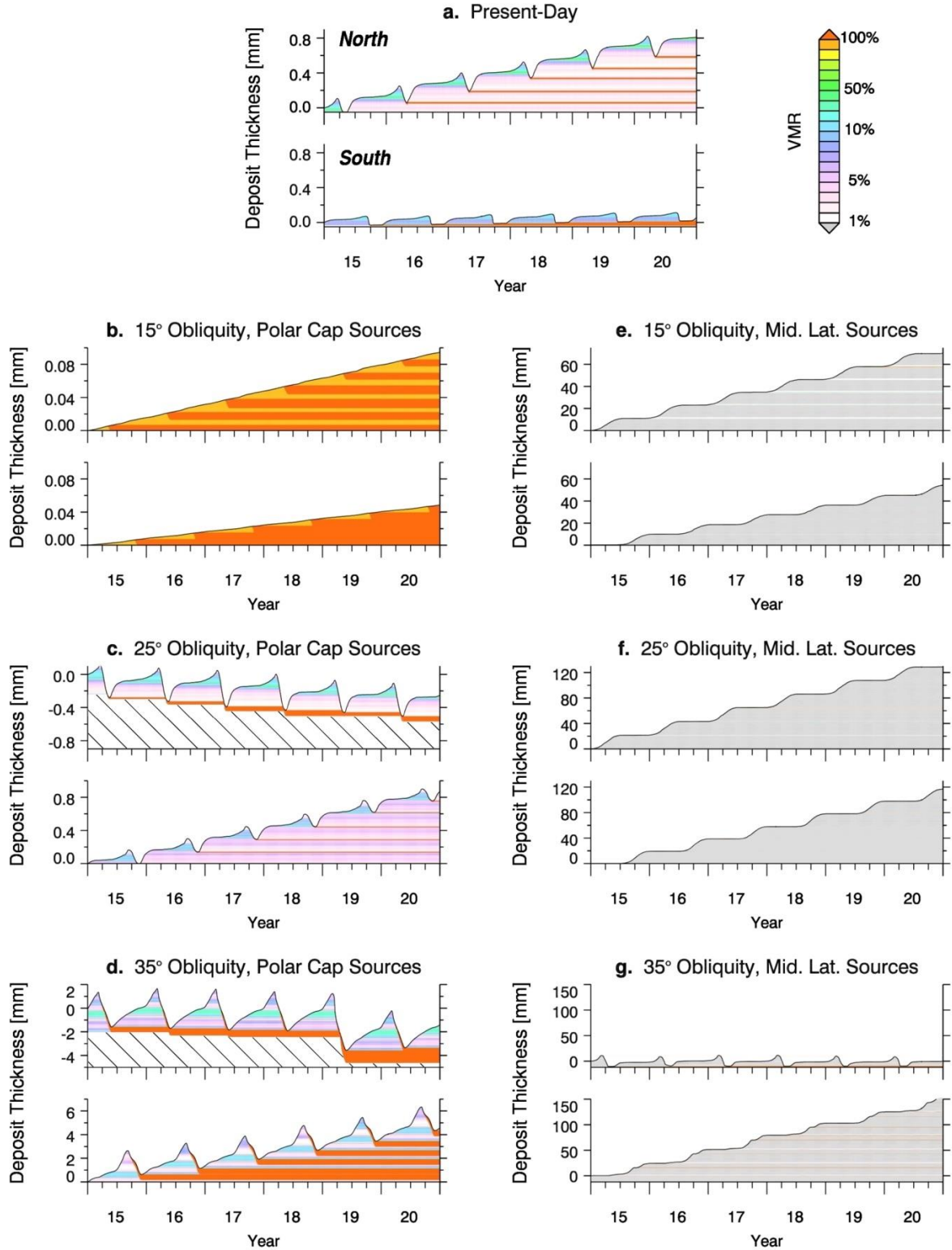

**Fig. 3. Multi-year evolution of the polar surface deposits of water ice and dust under the source and obliquity conditions described in section 3.** Colors represent the dust-to-water ice VMR of material accumulated/removed in ~5 sol increments. Gray represents VMR's < 1%. Red

represents dust lag deposits (100% dust). Hatched white areas represent prescribed source material containing 10% dust by volume.

Results are subsequently described for a set of simulations that employ circular orbits, obliquities ranging from 15° - 35°, and two different initial surface water ice distributions:

1) Symmetric polar water ice caps (poleward of 77.5° North and South), representing a period following a low obliquity epoch (~15°) that was perhaps characterized by the formation of such polar caps (Mischna et al, 2003; Levrard et al., 2007). The prescribed boundaries of the polar regions (77.5° North and South) also define the latitudinal extent of the prescribed surface water ice sources. This prescription prevents both simulated polar regions from simultaneously undergoing net annual accumulation of water ice, but allows for the annual deposition of water outside of the polar regions, transfer of water between the polar regions, or redistribution of water ice within the defined polar region boundaries. Polar cap source simulations prominently exhibit a net annual pole-to-pole transfer of water ice (Fig. 3b-d).

2) 5°-wide planet-encircling 'bands' of water ice at middle latitudes (centered at 40° North and South, encompassing all longitudes for simplicity), representing a period following a moderately-high (~35°) obliquity epoch that was perhaps characterized by the formation of glaciers at middle latitudes (Mischna et al., 2003; Levrard et al., 2007; Madeleine et al, 2009, 2014). The initially prescribed surface water ice sources are located within two zonal bands spanning 37.5° - 42.5° N/S. These midlatitude locations provide an inexhaustible source of water ice that is available to be deposited upon the polar surfaces (and other locations). Indeed, most midlatitude surface water ice source simulations predict a net annual transfer of water ice from the middle latitude source locations to the polar regions, driving net annual accumulation of water ice in both the north and south (Table 1c; Fig 3e-g).

The 25° obliquity simulations are described first for comparison with the present-day

(25.19°) simulation results. These represent intermediate obliquity/climate conditions among the

MGCM simulations that share their respective initial surface water ice source prescriptions.

## 3.2    25° Obliquity, Zero Eccentricity Conditions

### 3.2.1 Polar Surface Ice Sources / 25° Obliquity

The north polar region experiences a net annual loss of source and seasonally-deposited water ice of ~0.035 mm/yr, while the south polar region experiences a net annual gain of water ice of ~0.023 mm/yr. This differs from the present-day simulation which did not exhibit a net transfer of water ice from the north (source) to the south polar region. Yearly-integrated annual water ice deposition is almost entirely driven by snowfall in both polar regions (~85%) rather than by frost, a result that is ubiquitous among all polar cap source scenarios.

The simulated north and south polar regions accumulate dust at ~0.01 mm/yr and ~0.006 mm/yr, respectively, rates comparable to those predicted under present-day conditions (Table 1a,b). This simulation exhibits a seasonal evolution of polar dust deposition similar to that of the present-day simulation, despite exhibiting comparatively lower atmospheric dust abundances at southern midlatitudes in southern summer due to the circularization of the orbit, which suppresses dust storm activity. Annually-accumulated dust is dominated by dust cores, accounting for ~85% of deposited in the north and ~65% in the south, by volume.

The north polar surface annually accumulates a ~0.01 mm-thick dust lag deposit. The south polar surface accumulates an annual ~0.03 mm-thick water ice and dust "layer" with a dust-to-water ice VMR of ~22% (Fig. 3b). Year-end bulk polar VMR's of the north (100% dust lag) and south (22% mixture) are significantly greater than the polar regions' respective average snow particle VMRs of ~3.4% and ~2.5%.

### 3.2.2 Midlatitude Surface Ice Sources / 25° Obliquity

The north and south polar regions accumulate water ice at ~11 mm/yr and ~10 mm/yr, respectively, rates that are more than an order of magnitude greater than the inferred long-term average growth rate of the NPLD (Laskar et al., 2002; Levrard et al., 2007; Hvidberg et al., 2012). Yearly-integrated annual polar water ice deposition is overwhelmingly (> 99%) driven by snowfall in both polar regions rather than by frost, a result that is ubiquitous among most midlatitude source scenarios.

The north and south polar regions accumulate dust at rates of ~0.013 mm/yr and ~0.004 mm/yr, respectively. This simulation (and most described midlatitude source simulations) accumulate polar dust deposits that are overwhelmingly (> 99%) dominated by dust cores, in contrast to the present-day and polar cap source simulations which exhibit core-to-free dust ratios of ~55 - 85%. This results from the fact that midlatitude source simulations exhibit a comparatively greater degree of water ice-on-dust scavenging, i.e., a transition of free dust to core dust when water condenses on suspended dust particles. This is evident in a generally lower contribution of free dust to zonally-averaged atmospheric dust abundances (discussed in section 4.1.3).

The north polar surface accumulates an annual layer that is ~11 cm thick with a dust-to-water ice VMR of ~0.1% by volume. The simulated south polar surface accumulates an annual layer that is ~10 cm thick with a dust-to-water ice VMR of ~0.04%. The polar surface deposits exhibit rapid accumulation of dust-poor water ice from late local spring through late local summer and a lack of net deflation at the time of local summer solstice (Fig. 3f). Some summer solstice sublimation does occur but it is insufficient to dominate accumulation - this has the effect of only slightly (10%) raising the bulk VMRs of the north and south annual polar deposits (~0.1% and

~0.04%, respectively) relative to their respective average snow particle VMRs. Most of the following described midatitude source simulations also exhibit a small discrepancy between their average snow particle and bulk polar deposit VMRs (owing to lack of summer solstice net water ice removal) compared to the present-day and polar cap source simulations.

### 3.3  15° Obliquity, Zero Eccentricity Conditions

### 3.3.1  Polar Surface Ice Sources / 15° Obliquity

The north polar surface experiences a net annual accumulation of water ice of ~0.0001 mm/yr, while the south polar region experiences a net annual loss of ~0.0002 mm/yr, two orders of magnitude smaller than in the present-day and 25° obliquity polar surface water source simulations (Table 1a,b). The annual pole-to-pole transfer of water ice, here south-to-north (Fig. 3.8a), is opposite to the 25° obliquity polar cap source scenario (Fig. 3c).

The north and south polar regions annually accumulate dust at ~0.008 mm/yr and ~0.004 mm/yr, respectively. These rates are approximately an order of magnitude smaller than predicted under present-day and 25° obliquity/zero eccentricity polar cap source conditions (Table 1a,b) and are consistent with decreased atmospheric dust loading with decreasing obliquity under polar cap source conditions (discussed in section 4.1.2). Deposited dust is approximately evenly-distributed between core (~55%) and free dust in both polar regions, in contrast with the 25° obliquity polar cap source scenario in which core dust deposition dominates.

The reduced water ice deposition rate but comparable dust deposition rate relative to the 25° obliquity scenario results in a much greater dust-to-water ice VMR of north polar deposited materials. The north polar surface annually accumulates a ~0.008 mm-thick layer with a dust-to-water ice VMR of 98%, while the south polar surface annually accumulates a ~0.004 mm-thick dust lag deposit (Fig. 3b). Year-end bulk polar VMR's of the north (98% mixture) and south (100% dust lag deposit) are only slightly larger than the polar regions' respective average snow particle VMRs of ~95% and ~85%.

### 3.3.2 Midlatitude Surface Ice Sources / 15° Obliquity

The north and south polar regions annually accumulate surface water ice at rates of ~6.2 and ~4.8 mm/yr, respectively. These rates are an order of magnitude greater than the inferred long-term average growth rate of the NPLD (Laskar et al., 2002; Levrard et al., 2007; Hvidberg et al., 2012) but are ~half of the rate predicted under 25° obliquity/zero eccentricity midlatitude source conditions (Table 1c). This is consistent with increased atmospheric water abundances with increasing obliquity under midlatitude source conditions (discussed in section 4.1.3).

The north and south polar regions annually accumulate dust at rates of ~0.028 mm/yr and ~0.008 mm/yr, respectively. The polar dust deposits accumulated in this scenario are overwhelmingly (> 99%) dominated by core dust, consistent with the results of other described midlatitude source simulations.

The north polar surface accumulates a ~6.2 mm-thick annual layer with a dust-to-water ice VMR of ~0.45%, while the simulated south polar surface accumulates a ~4.8 mm-thick annual layer with a dust-to-water ice VMR of ~0.15% (Fig. 3e). Summertime sublimation, though insufficient to drive net instantaneous removal of water ice, increases the polar annual deposits' bulk VMRs by ~10% relative to the north and south's respective average snow particle VMRs of 0.40% and 0.14%.

## 3.4    35° Obliquity, Zero Eccentricity Conditions

### 3.4.1 Polar Surface Ice Sources / 35° Obliquity

The north polar surface experiences a net annual loss of source and seasonally-deposited water ice at a rate of ~0.44 mm/yr, while the south polar region experiences a net annual accumulation of water ice at ~0.14 mm/yr. This simulation resembles the present-day and the 25° obliquity polar surface water source simulations in that it exhibits a north-to-south annual pole-to-pole transfer of water ice (Fig. 3d), but with order of magnitude greater annual polar water ice accumulation and loss rates (Table 1a,b). This reflects increased atmospheric water abundances with increased obliquity under polar cap source conditions (discussed in section 4.1.2).

Both the north and south polar regions annually accumulate dust at ~0.25 mm/yr, an order of magnitude greater than under present-day and 25° obliquity/zero eccentricity polar cap source conditions (Table 1a,b), resulting from an order of magnitude enhancement in atmospheric dust loading at high latitudes during both hemispheres' local summer that accounts for most polar annual dust deposition (discussed in section 4.1.2). Annually-accumulated dust is dominated by free (rather than core) dust in the north (free dust is ~70% of deposited dust) and south (free dust is ~85% of deposited dust). This contrasts with the 25° obliquity polar cap source simulation which formed polar surface dust deposits primarily consisting of core dust. It also contrasts with the 15° obliquity polar cap source scenario which formed polar surface dust deposits consisting of an approximately even distribution of core and free dust in both polar regions.

The north polar surface annually accumulates a ~0.25 cm dust lag deposit, while the south polar surface annually accumulates a ~0.4 mm-thick layer with a dust-to-water ice VMR of ~70% (Fig. 3d). The net annual removal of water ice (north), and the summertime retention of previously-deposited core dust and deposition of free dust (both poles) produces year-end bulk polar VMR's of the north (100% dust lag) and south (70%) that are significantly greater than the polar regions' respective average snow particle VMRs of ~5% and ~4%.

### 3.4.2 Midlatitude Surface Ice Sources / 35° Obliquity

The north polar surface undergoes no annual water ice accumulation, while the south polar region annually accumulates water ice at a rate of ~14 mm/yr. The south's annual water ice accumulation rate is more than an order of magnitude greater than the inferred ~0.5 mm/yr long-term average growth rate of the NPLD (Laskar et al., 2002; Levrard et al., 2007; Hvidberg et al., 2012) and is ~25% greater than predicted under 25° obliquity/zero eccentricity midlatitude source conditions (Table 1c). This is consistent with increased atmospheric water abundances with increasing obliquity under midlatitude source conditions (discussed in section 4.1.3).

The north and south polar regions accumulate dust at rates of ~0.08 mm/yr and ~0.01 mm/yr, respectively, an order of magnitude larger in the north and ~30% larger in the south compared to the 25° obliquity middle latitude source simulation. This is consistent with enhanced atmospheric dust loading at high northern latitudes during its early northern summer but the lack of a corresponding event in the south (discussed in section 4.1.3). The polar dust deposits are overwhelmingly dominated by dust cores, consistent with the results of other described middle latitude source simulations.

The north polar surface annually accumulates a ~0.08 mm-thick sublimation dust lag deposit, while the south polar surface accumulates a ~14 mm-thick annual layer with a dust-to-water ice VMR of ~0.1% (Fig. 3g). In the north, the exhaustion of the seasonal water ice deposit, and the retention of previously-deposited core dust subsequent to sublimation and the deposition of free dust, annually produces a dust lag deposit. In the south, sublimation, which briefly drives net water ice removal and dust lag formation in summer, is sufficient to raise the bulk VMR of the annual deposit by a factor of five relative to the average VMR of precipitated snow particles (~0.02%).

## 4. Discussion

In this section, we assess the dependency of net annual polar deposition rates on specified obliquity/source location conditions, and the climate-related factors which drive those rates. The per-year deposition rates arising from the MGCM simulation results are temporally insufficient to investigate long-term PLD depositional processes over orbital timescales. We therefore introduce an "integration model" which employs the MGCM deposition rates to assess the vertical structure of polar layered deposits arising from the obliquity/water source-dependent deposition rates over five obliquity cycles.

### 4.1 Comparison of Paleoclimate Results

### 4.1.1 Annual Rates of Water Ice and Dust Deposition at the Poles

The polar deposition results of the simulated scenarios explored in this work exhibit a range of net annual polar accumulation rates of water ice (spanning -1 mm/yr to +14 mm/yr) and dust (spanning +0.003 mm/yr to +0.3 mm/yr). Accounting for the nucleation of water ice on dust enables polar dust accumulation rates to be decomposed into that of dust cores (+0.0018 mm/yr to +0.11 mm/yr) and into that of free dust (-0.19 mm/yr to +0.22 mm/yr).

The predicted range of water ice accumulation rates (Table 1; Fig. 4a,d) demonstrate that the polar regions can undergo either net annual accumulation (with or without polar water ice cap sources) or removal (from polar cap sources) of surface water ice when obliquity is varied between extrema values characteristic of Mars' recent orbital history (present to ~5 mya) under zero eccentricity conditions. Predicted net annual polar water ice deposition rates (spanning -1 mm/yr to +14 mm/yr) fall within the range of inferred rates (spanning -4 to +30 mm/yr) from estimated NPLD impact crater infill within the past ~100 kyrs (Herkenhoff and Plaut, 2000;

Banks et al., 2010; Landis et al., 2017) and from model-based investigations that invoked comparable obliquity and eccentricity conditions (Levrard et al., 2007; Mischna et al., 2003; Hvidberg et al., 2012).

Net annual total (core + free) dust deposition is positive in both polar regions in all tested scenarios, implying continuous but obliquity cycle timescale-varying polar dust accumulation over the build-up history of the PLD (Table 1; Fig. 4b,e). Predicted net annual polar deposition rates of total (core + free) dust (~ +0.003 mm/yr to +0.3 mm/yr) encompass rates predicted by Hvidberg et al. (2012) (~0.01 mm/yr to ~0.02 mm/yr), who used an insolation-based parameterized model that did not invoke prescribed surface water ice or dust source locations. Predicted net annual polar deposition rates of total dust (~ +0.003 mm/yr to +0.3 mm/yr) encompass typical rates of ~ +0.007 mm/yr predicted by Newman et al. (2005) over the same obliquity range. Simulated present-day conditions predict a NPLD resurfacing rate of ~ +0.1 mm/yr that is an order of magnitude smaller than the ~ +0.7 mm/yr rate predicted by Hvidberg et al. (2012) for the last 1000 years. Accounting for imprecise water and dust conservation does not increase reported rates by more than an order of magnitude (as discussed in section 2), and does not affect the described qualitative comparisons between these and previously estimated rates.

**Table 1. Sensitivity of annual net polar water and dust deposition to various source and obliquity conditions.** Values are reported in mm/Earth year for **a.** present-day, **b.** polar cap source, and **c.** midlatitude sources (these conditions were discussed in section 3. This table includes results for all five tested obliquity values: 15°, 20°, 25°, 30° and 35°). Results for all 56 MGCM simulations are contained in **Table S1** (supplementary).

**a.**

| | Present-Day (North Residual Cap Sources) | | | | | |
|---|---|---|---|---|---|---|
| | North Polar Region | | | South Polar Region | | |
| Obliquity | Water [mm/yr] | Dust [mm/yr] | VMR [%] | Water [mm/yr] | Dust [mm/yr] | VMR [%] |
| 25.19° | 0.063 | 0.011 | 14 | 0 | 0.005 | 100 |

**b.**

| | North and South Polar Cap Sources (77.5° – 87.5° North/South) | | | | | |
|---|---|---|---|---|---|---|
| | North Polar Region | | | South Polar Region | | |
| Obliquity | Water [mm/yr] | Dust [mm/yr] | VMR [%] | Water [mm/yr] | Dust [mm/yr] | VMR [%] |
| 15° | 0.0001 | 0.008 | 98 | -0.0002 | 0.004 | 100 |
| 20° | -0.002 | 0.009 | 100 | 0.0008 | 0.005 | 86 |
| 25° | -0.035 | 0.010 | 100 | 0.023 | 0.006 | 22 |
| 30° | 1.1 | 0.031 | 3 | -1.0 | 0.021 | 100 |
| 35° | -0.44 | 0.25 | 100 | 0.14 | 0.27 | 67 |

**c.**

| | Midlatitude Sources (37.5° – 42.5° North/South) | | | | | |
|---|---|---|---|---|---|---|
| | North Polar Region | | | South Polar Region | | |
| Obliquity | Water [mm/yr] | Dust [mm/yr] | VMR [%] | Water [mm/yr] | Dust [mm/yr] | VMR [%] |
| 15° | 6.2 | 0.028 | 0.45 | 4.8 | 0.008 | 0.16 |
| 20° | 8.7 | 0.020 | 0.22 | 7.4 | 0.006 | 0.09 |
| 25° | 11 | 0.013 | 0.11 | 10 | 0.004 | 0.04 |
| 30° | 13 | 0.010 | 0.08 | 13 | 0.003 | 0.02 |
| 35° | 0 | 0.082 | 100 | 14 | 0.01 | 0.07 |

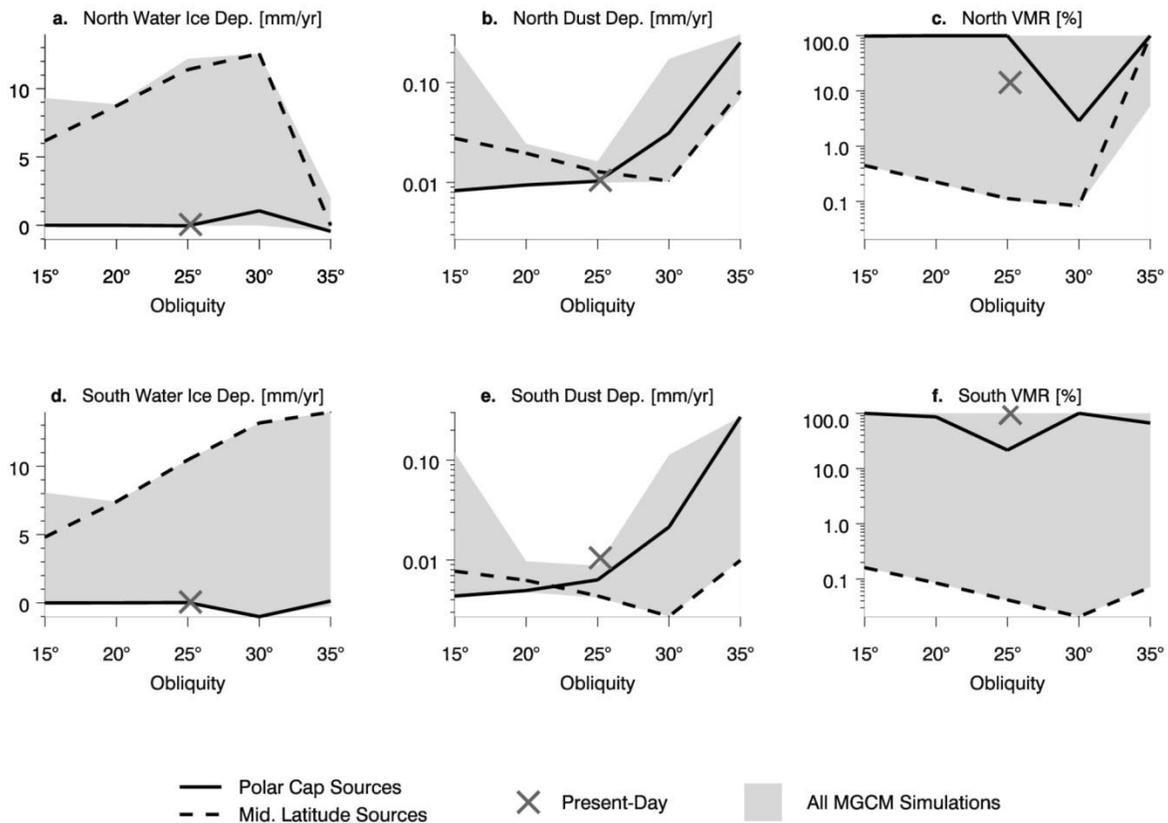

**Fig. 4. Sensitivity of annual net polar water and dust deposition rates in mm/Earth year to obliquity variations.** The solid curve represents "polar cap source" simulation results. The dashed curve represents "midlatitude" source simulations results. The X symbol represents present-day simulation results. The gray filled region represents the full range of results for all 56 MGCM simulations. **a.** annual water ice deposition rates in the north polar region vs. obliquity **b.** annual dust deposition rates in the north polar region vs. obliquity **c.** north polar dust-to-water ice VMR calculated from a. and b. **d.-f.** Same as (a-c), but for the south polar region.

The results presented in Table 1 and Fig. 4 suggest that two main types of layers may be annually produced in either polar region over the full range of tested configurations 1) dust-water ice mixtures comprised of between ~0.01 % to 98 % dust by volume, or 2) sublimation dust lag deposits (100% dust).

When the polar deposits are undergoing their most rapid seasonal growth in local spring and late local summer (when snowfall rates are maximized), the dust-to-ice mixing ratio of material actively accumulating on the polar surface is comparable to that of precipitating snow in the overlying atmosphere ($\leq 0.1\%$). Because the dust-to-water ice VMR of precipitated snow varies during the year, seasonal variations in dust deposition (maximizing around the equinoxes at low obliquities or during local summer at high obliquities) do not directly correlate with those of snowfall. Summertime sublimation ultimately increases the bulk dust-to-water ice VMR of the annual deposits relative to precipitated snow particles. Nevertheless, the microphysical coupling of the dust and water cycles via nucleation and their co-deposition via snowfall provides the primary means by which water and dust are delivered to the polar surfaces, making such coupling an important consideration for model-based investigations of PLD layer formation processes.

### 4.1.2 Climate Conditions that Drive the Results of the "Polar Cap Source" Simulations

Zero eccentricity simulations that employ simultaneous north and south polar ice caps generally predict an annual polar cap-to-cap exchange of water ice (Table 1b; Fig. 3b-d). A dust lag deposit forms in the net subliming polar region, while a dust-water ice mixture forms in the net accumulating polar region. Increasing obliquity increases polar summer temperatures and the amount of water involved in the seasonal water cycle (Fig. 5b-d) as predicted by previous investigations (Richardson and Wilson, 2002; Mischna et al., 2003; Levrard et al., 2004; Forget et al., 2006), which here results in an increase in the amount of water ice that is annually exchanged between the prescribed polar caps. When the north polar region is the subliming pole (obliquities of 20°, 25°, and 35°), higher local summer solstice surface temperatures in the north polar region produce greater instantaneous sublimation rates in summer compared to the south, resulting in annual loss of water ice in the north and accumulation in the south. The 15° and 30° obliquity scenarios exhibit instead a south-to-north annual exchange of water ice. At 30° the exchange is due to a recurrent southern summer dust event that enhances total downward radiative flux on the south cap, inducing greater sublimation in that season. At 15° the exchange is due to differences in the retreat times of the north and south perennial $CO_2$ ice cap edges (and the exposure of water ice along their peripheries). Both polar regions exhibit comparable late springtime rates of $CO_2$ sublimation, but the periphery of the north $CO_2$ cap where water ice is seasonally exposed (77.5° - 82.5° N) is ~ 6% / 4 cm thicker before the onset of its sublimation and exhibits a smaller sublimation rate around local summer solstice compared to the south, therefore requiring an additional ~60° of $L_s$ in order to expose underlying water ice. The north builds a thicker $CO_2$ ice cap in the preceding local spring because higher surface pressure conditions at the north polar surface initiate $CO_2$ condensation at warmer temperatures (15° of $L_s$ earlier in local spring compared to the south). The 35° obliquity polar cap source scenario demonstrates that unique climate conditions can arise inter-annually under a fixed obliquity/source configuration. For instance, year 19 exhibits substantially greater northern summertime sublimation compared to surrounding years (Fig. 3d), arising from reduced northern midlatitude atmospheric dust and corresponding increased downward radiative flux enhancing surface water ice sublimation.

The polar surfaces accumulate dust at all obliquities (Table 1b; Fig. 4b,e), primarily as dust cores nucleated by snow particles. Most dust deposited on the polar surfaces is supplied by surface wind stress dust lifting along the growing/receding seasonal ice caps in early-mid spring and autumn. Polar dust deposition increases with increased obliquity due to increased surface wind stress lifting, arising from increased temperature gradients and induced surface wind stress magnitudes along the more latitudinally-extensive seasonal $CO_2$ cap edge. This is reflected in increased atmospheric dust abundances with increased obliquity under these polar cap source conditions (Fig. 6b-d). The 35° obliquity simulation exhibits even greater annual polar dust deposition rates, due to the development of additional surface wind stress dust lifting events that occur at northern and southern midlatitudes at their respective local summer solstices, increasing atmospheric dust abundances globally (Fig. 6d).

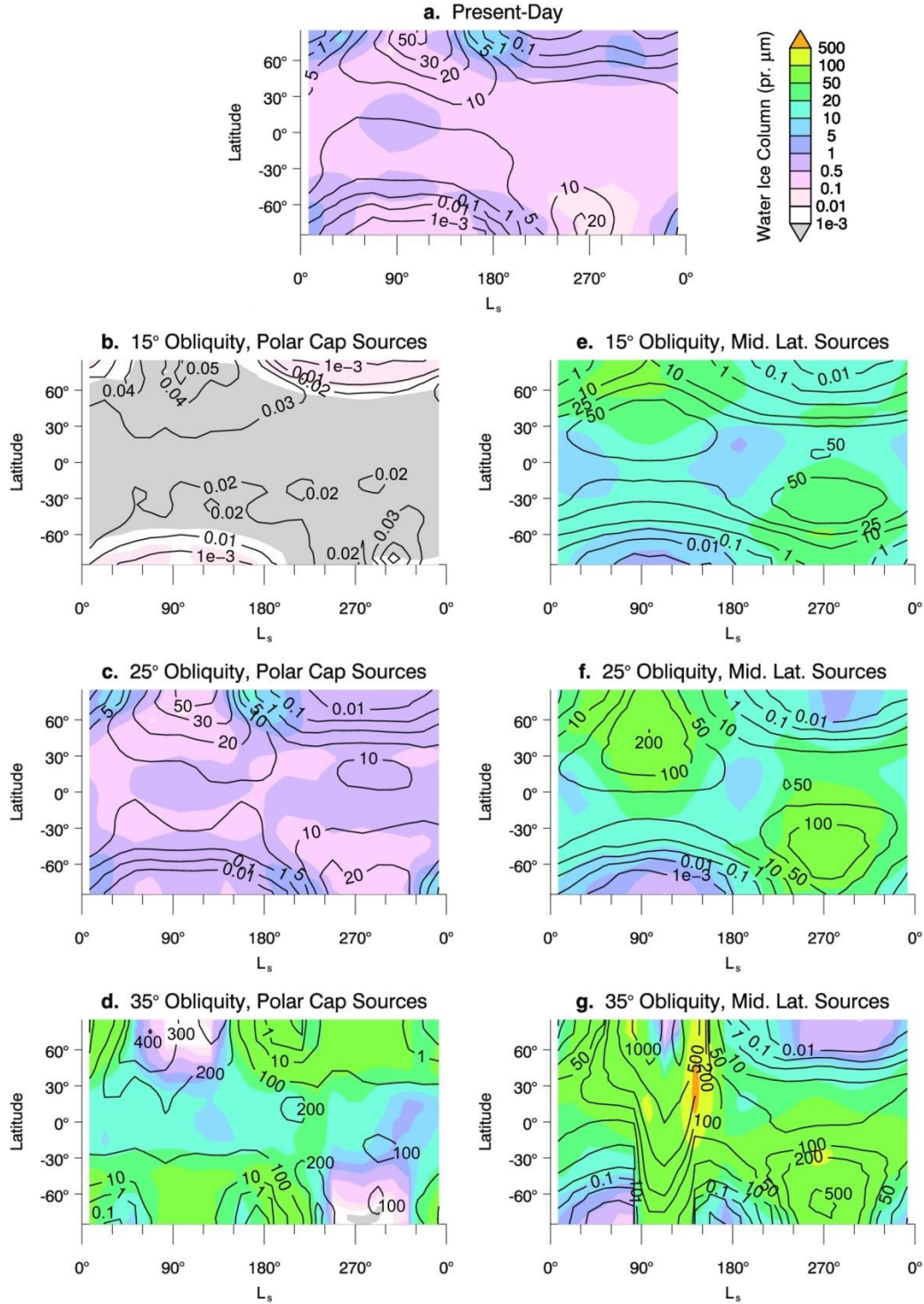

**Fig. 5. Zonal evolution of the water ice column (colored contours, pr. μm) and water vapor column (over-plotted contours, pr. μm) as a function of time of year, under various source and obliquity conditions. a.** present-day, **b-d.** polar cap source, and **e-g.** midlatitude source conditions (the conditions discussed in section 3). Note that the water vapor column contour intervals differ between plots.

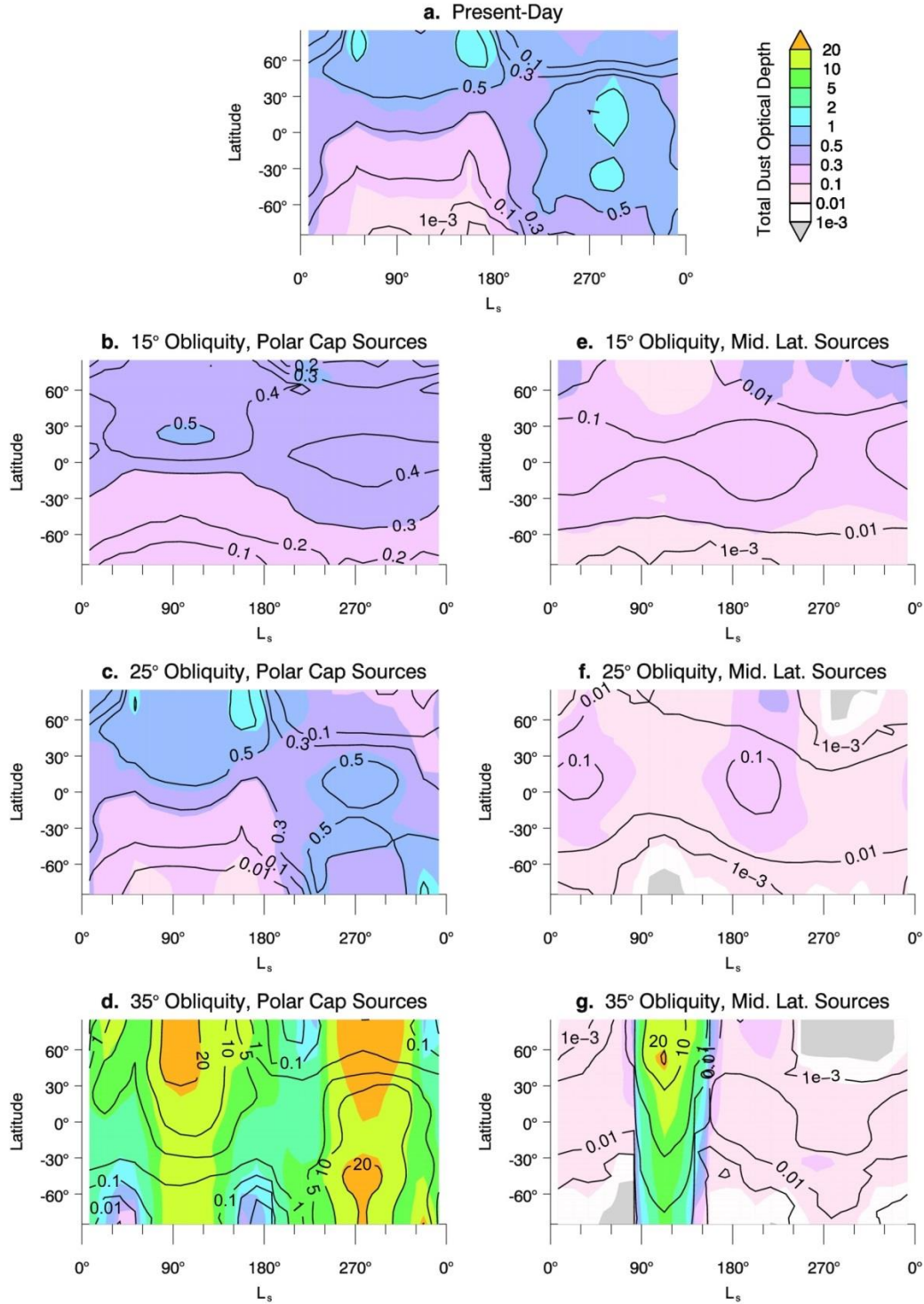

**Fig. 6. Zonal evolution of the total dust optical depth (colored contours) and free dust optical depth (over-plotted contours) as a function of time of year, under various source and obliquity conditions. a.** present-day, **b-d.** polar cap source, and **e-g.** midlatitude source conditions (the conditions discussed in section 3). Note that the free dust optical depth contour intervals differ between plots.

### 4.1.3 Climate Conditions Driving the Results of "Midlatitude Source" Simulations

Zero eccentricity simulations that employ simultaneous north and south midlatitude glacial deposits generally predict rapid annual polar water ice accumulation at the expense of the midlatitude ice (Table 1c; Fig. 3e-g; Fig. 4a,d). Polar accumulation (and midlatitude loss) of water ice increases with increased obliquity at rates that are typically more than an order of magnitude greater than the inferred accumulation rate of the NPLD (0.5 mm/yr) (Laskar et al., 2002; Levrard et al., 2007; Hvidberg et al., 2012), reflecting relatively high atmospheric water abundances that increase with increasing obliquity (Fig. 5e-g).

The polar surfaces accumulate dust at all obliquities (Table 1c; Fig. 4b,e). However, unlike the polar cap source simulation results, polar dust deposition rates decrease with increasing obliquity. Increasing overlap between the extensive seasonal $CO_2$ caps and the midlatitude surface water ice deposits reduces cap-edge surface wind stress lifting. This is reflected in a general decrease in zonally-averaged dust abundances with increased obliquity (Fig. 6e-g). The 35° obliquity simulation exhibits greater annual polar dust deposition compared to lower obliquities due to the addition of a northern summer, northern high latitude-centered dust storm that globally enhances atmospheric dust loading (Fig. 6g). Though this event simultaneously enhances dust deposition on the south cap, atmospheric dust abundances are maximized at high northern latitudes, resulting in order of magnitude greater deposited dust in the north polar region compared to the south. Core dust comprises > 99% of annually deposited dust, resulting from greater rates of scavenging of water ice on dust in the midlatitude source simulations compared to polar cap source simulations at a given obliquity, as reflected by free dust globally contributing a smaller fraction to zonally-averaged atmospheric dust abundances (Fig. 6e-g) compared to polar cap source simulations (Fig. 6b-d).

## 4.2 A Model of PLD Growth and Layer Formation at Low-Eccentricity

The MGCM results indicate that the annual rates of water ice and dust deposition exhibit a complex dependence on both obliquity and the global distribution of surface water ice (Table 1; Fig. 4). The range of predicted polar deposition rates are generally consistent with those inferred by previous investigations (Herkenhoff and Plaut, 2000; Mischna et al., 2003; Newman et al., 2005; Levrard et al., 2007; Banks et al., 2010; Hvidberg et al., 2012; Landis et al., 2017). Analysis of GCM depositional processes reveal the potential importance of the microphysical coupling of the water and dust cycles (nucleation and snowfall) on the formation of polar surface layers over multiannual timescales. However, the MGCM simulations conducted and presented here span only ~20 martian years, not nearly long enough to assess resultant PLD effects over the $10^4$ - $10^5$ year time scales relevant for the formation of the PLDs' observable stratigraphy (Plaut et al., 1988; Herkenhoff et al., 2000; Laskar et al., 2002; Levrard et al., 2007; Milkovich and Head, 2005; Fishbaugh et al., 2010a; Hvidberg et al., 2012; Becerra et al., 2017). Similarly to Levrard et al. (2007), we investigate the formation history of the NPLD over multi-obliquity cycle timescales, using deposition rates from the suite of MGCM simulations described in section 3. These deposition rates were ingested into a time-marching model that integrates them over a 700 kyr-spanning low eccentricity epoch starting 1.7 Ma.

MGCM-derived annual polar accumulation or loss rates are zonally averaged in 5° intervals between 87.5° S and 87.5° N (the greatest poleward extent of the GCM), and are integrated through time as a means of tracking the global redistribution of water ice and dust. The latter includes both 'free' and ice-nucleated 'core' dust.

The algorithm for tracking the subsequent evolution of surface water ice and dust reservoirs in the integration model proceeds as follows:

1) At time t = 0 (~1.7 mya) and at its corresponding obliquity (Fig. 1), emplace an initial quantity of surface water ice at prescribed latitudes representing the initial distribution of surface water ice reservoirs.

2) Dynamically identify "best match" MGCM simulations that encompass the obliquity and surface water ice source specified above.

3) Interpolate among best match MGCM simulations at the specified obliquity to determine applicable deposition rates.

4) Propagate forward the deposition rates, employing 1000 year time steps (the highest time resolution of the reference obliquity history (Laskar et al., 2004).

5) Repeat steps 2 – 4

At the model's polar latitudes, discrete changes in surface water ice and dust thickness are summed to define the time-evolving thickness of the simulated polar deposit. The deposit's vertical structure (VMR vs height) is calculated from the relative quantities of water ice and dust deposited on the polar surfaces during each 1000-year time-step. During periods of net water ice removal, the integration model accounts for dust lag formation by retaining dust that was contained within removed layers, and placing that dust on the deposit's surface.

The following eleven initial integration model water ice reservoir distributions are tested:

1. Simultaneous north and south polar caps (surface water ice poleward of 77.5° N/S)

2. A north polar cap (surface water ice northward of 77.5° N) only

3. A south polar cap (surface water ice southward of 77.5° S) only

4. Simultaneous north and south midlatitude sources (37.5° – 42.5° N/S)

5. Simultaneous north and south high latitude sources (57.5° – 62.5° N/S)

6. A northern midlatitudee source (37.5° – 42.5° N) only

7. A northern high latitude source (57.5° – 62.5° N) only

8. A southern midlatitude source (37.5° – 42.5° S) only

9. A southern high latitude source (57.5° – 62.5° S) only

10. Simultaneous north and south polar caps and simultaneous north and south midlatitude sources (surface water ice poleward of 77.5° N/S and 37.5° – 42.5° N/S)

11. Simultaneous north and south polar caps and simultaneous north and south high latitude sources (surface water ice poleward of 77.5° N/S and 57.5° – 62.5° N/S)

Under each tested water ice reservoir distribution, a variety of initial quantities of water ice were prescribed within those reservoirs (ranging from 100 to 2500 m-thick), and a variety of "threshold" thicknesses are specified (ranging from 1 micron to 100 m-thick) that a prescribed or deposited ice deposit must exceed to be considered a water ice source at a given time step. Threshold surface water ice thickness values are specified to prevent the sublimation of more water than is available from surface sources when GCM-derived rates are propagated forward in 1000 year time steps.

Integration model results indicate that, except during the first obliquity cycle, there are no long term implications for the created PLD structure that depend on the initial obliquity condition specified.

Fig. 7d displays an example of the vertical structure characteristic of the formed northern surface deposits arising from the integration model. The integration model simulation highlighted here is initialized at an obliquity cycle minimum (~15° obliquity at 1.68 mya) with initial 1500-m thick polar ice caps. In the north, two dust-rich layers are formed per obliquity cycle (Fig. 7d): (1) a ~30 m-thick layer with a typical dust-to-water ice VMR of ~20 - 30% that forms at high obliquity (~30°), and (2) a 0.5 m-thick dust lag deposit which forms at low obliquity (~20°). The ~ 30 m layers may correspond to a prominent ~30 m wavelength observed in brightness and topographic profiles of the upper ~300 m of the NPLD (Milkovich and Head, 2005; Fishbaugh et al., 2010a; Becerra et al., 2017), which would imply that this measured wavelength is linked to ~120 kyr obliquity variations. The second dust-rich layer type would be undetectable via current radar imaging capabilities as it exhibits a thickness that is more than an order of magnitude lower than SHARAD's ~10 m vertical resolution limit, but its thickness is comparable to the ~decimeter resolution of HiRISE. This thickness is only a factor of ~2 smaller than the thinnest layers observed in the PLD (termed "thin layers"), which occur in tightly-packed bundles between marker bed deposits (Fishbaugh et al., 2010a). Dust-rich layer pairs bound a ~10 m-thick layer with a relatively-low dust-to-water ice VMR of ~3% (Fig. 7d), a ~10 m-thick layer being within HiRISE and SHARAD detection capabilities (McEwen et al., 2007; Phillips et al., 2008). To supply the long-term accumulation of water ice in the north, the south polar cap deflates by hundreds of meters and accumulates a ~10's of meters-thick sublimation dust lag deposit on its surface (Fig. 7c). Growth of the sublimation dust lag might in reality be slowed or prevented by surface dust lifting if the dust remains exposed to the surface and isn't cemented by water ice, and this lag could have in turn enhanced/inhibited the subsequent sublimation of underlying water ice (thus modulating its own growth rate).

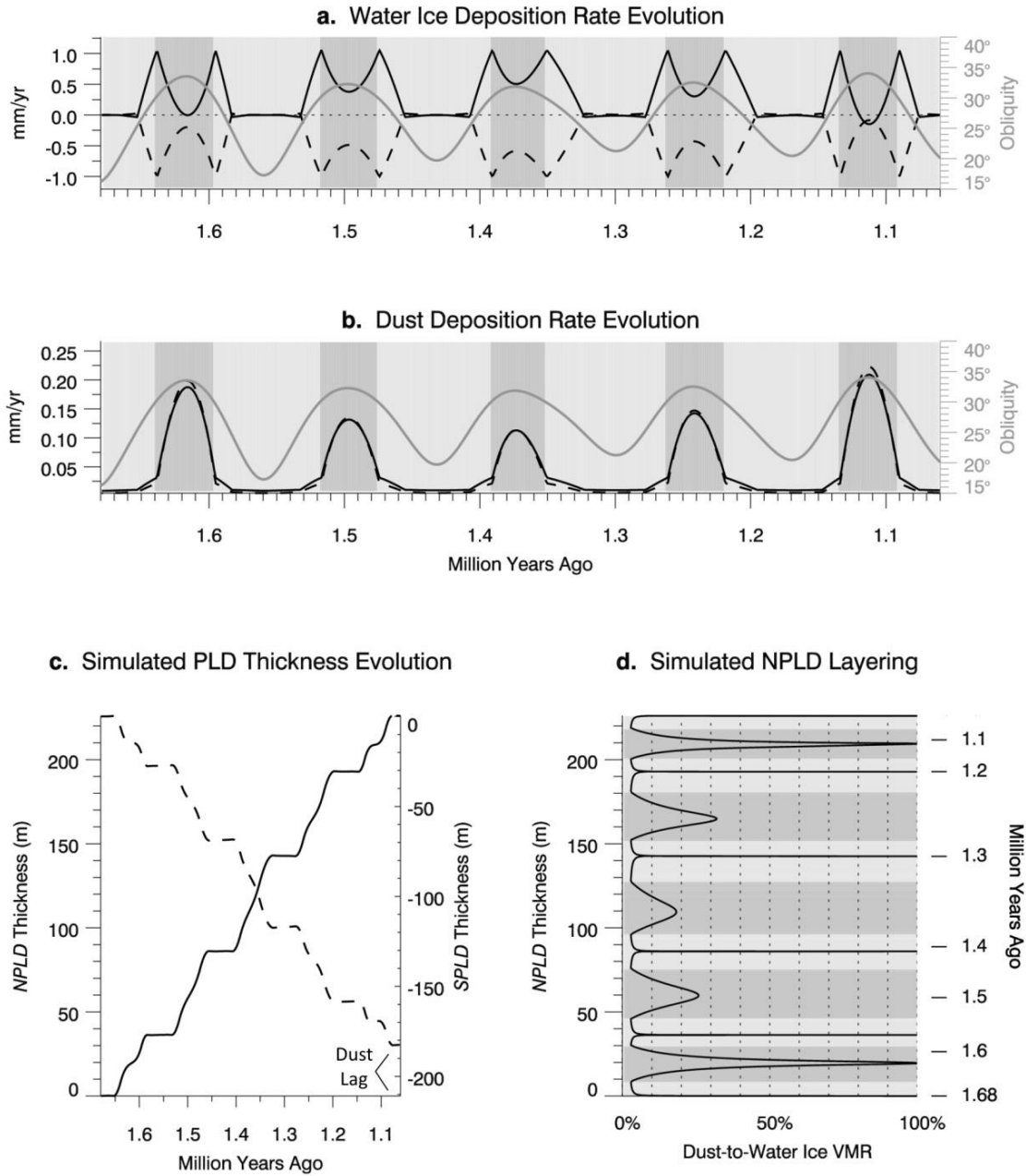

**Fig. 7. NPLD accumulation history from ~1.7 – 1.0 Ma, obtained with an integration model simulation that initially employs 1500 m-thick north and south polar water ice caps and accumulates a northern deposit at a time-averaged rate of ~0.36 mm/yr. a.** North (solid) and south (dashed) water ice net deposition rate vs time. Gray curves represent Mars' obliquity variations over the same period, demonstrating deposition-obliquity phasing. **b.** North (solid) and south (dashed) dust net deposition rate vs time. Line representation is the same as for a. **c.** NPLD (solid) and SPLD (dashed) thickness vs time. **d.** Fractional dust-to-water ice VMR vs. depth of the NPLD. Dark gray rectangles highlight thick dust-rich layers formed at high obliquity when water ice deposition and dust deposition are high. Light gray rectangles highlight dust lag deposits formed at low obliquity due to net water ice removal.

A parameter space comprised of 1782 total integration model simulations (162 integration model simulations for each of the 11 different initial surface water ice distributions) was tested. We focused on the 700 integration model simulations that accumulated a northern deposit at time-average rates of 0.1 - 1 mm/yr, close to the NPLD's inferred average growth rate of 0.5 mm/yr (Laskar et al., 2002; Levrard et al., 2007; Hvidberg et al., 2012). A number of different initial surface water ice distributions are represented among these simulations. 25% of them were initialized with north and south polar ice caps and retain those caps over the entire ~700 kyr integration period (resembling the simulation in Fig. 4). 40% were initialized with relatively thin north and/or south midlatitude deposits, but immediately begin forming north and south polar caps and exhaust their midlatitude deposits no later than the end of the second modelled obliquity cycle. Though these integration model simulations evolve towards a persistent north and south polar cap source configuration and thus ultimately produce a stratigraphy like that represented in Fig. 7d, the initial transfer of water from the midlatitude to the polar regions at ~cm/yr rates (Table 1b) first forms a ~100 m-thick, ~10% VMR deposit on both polar surfaces. An additional ~25% were initialized with a south polar ice cap only, which immediately begins undergoing deflation and supplies the ongoing accumulation of a northern ice cap (evolving again towards a south-to-north cap exchange configuration). Finally, the remaining 10% of the 700 models represented an unphysical scenario in which the north cap accumulated a ~70 m-thick water ice and dust mixture despite water being initially sourced from the north cap alone. This reflects the integration model's inability to perfectly account for water conservation.

These integration model simulation results suggest that a long term transfer of water from the south polar ice cap to the north polar cap could supply the a build-up of the NPLD at close to the inferred ~0.5 mm/yr rate (Laskar et al., 2002). This transfer would require a significant amount of water ice to have been accumulated in the south polar region prior to ~1.7 million years ago to sustain such a south-to-north exchange over the ~0.7 million year integration period investigated here, which is plausible given the SPLD's estimated age of 10 - 100 myr (Plaut et al., 1988; Herkenhoff et al., 2000). The question of whether the south cap's retention of water ice into the present day - despite significant ablation during the most recent low-eccentricity epoch - could have resulted from a once greater thickness of ice and/or periodic replenishment of its ice is beyond the scope of this work.

The simulated NPLD stratigraphy produced in this work can be compared to an NPLD stratigraphy simulated by Hvidberg et al. (2012), which expressed time-varying net deposition rates of water ice and dust as parameterized functions of insolation. That model accounted for both obliquity and eccentricity variations, but did not explicitly consider transport processes or variations in surface water ice source locations. Their model that best-fit an NPLD stratigraphic column of layers exposed within one trough was characterized by an inverse relationship between annual north pole water ice deposition rates and obliquity, which results in net deposition below an obliquity threshold of ~30$^\circ$ but net removal above this threshold. This relationship resulted from water ice deposition rates being computed as a sum of a north polar surface temperature-dependent sublimation term (directly proportional to obliquity) and a constant north polar deposition term. The parameterization of the dust deposition rate in that study was based on the calculated north pole-to-equator surface temperature gradient. An inverse relationship between annual north polar dust deposition rates and obliquity was predicted, perhaps consistent with increased low-latitude dust devil lifting with decreased obliquity (Newman et al., 2002).

Our integration model results and those of Hvidberg et al. (2012) both predict that maximum dust deposition rates occur in conjunction with maximum water ice deposition rates, and vice versa (Fig. 7a,b). However, the integration model predicts that maximum net annual deposition rates correspond to obliquity maxima, while Hvidberg et al. (2012) predict that maximum deposition rates occur in conjunction with obliquity minima. The minimum VMR of simulated layers produced by Hvidberg et al. (2012) and this work are comparable (~ a few percent dust by volume), but the dust-rich icy layers produced here (when obliquity is maximized) have VMRs that are a factor of 7 to 30 greater than those produced by Hvidberg et al. (2012) (when obliquity is minimized). These discrepancies likely result from considerable differences in how annual polar water ice and dust deposition have been modelled between these investigations, however there is significant qualitative similarity between the simulated NPLD layered structures and the formation mechanisms that produce them, supporting the previous suggestions that obliquity and orbital variations are important factors in producing the PLD vertical structure.

Our simulated NPLD stratigraphy can also be compared to observationally-constrained estimates for NPLD layer VMRs. Lalich et al. (2019) combined orbital radar reflectivity measurements with physical layer models to produce simulated VMR vs. depth profiles (to ~500 m-depth) for multiple study sites throughout the NPLD. These observationally-inferred profiles are most appropriately compared to our simulated dust-rich icy layers formed during low obliquity excursions, since 0.5 m-thick dust lag deposits fall below the vertical resolution of SHARAD. Integration-model produced dust-rich icy layers exhibit individual VMRs between ~20% - 30% and column-averaged VMRs between ~5% - 25%. While these column-averaged and individual layer VMRs are greater than the inferred bulk dust concentration of the entire

NPLD (~5%), Lalich et al. (2019) similarly predicts high column-averaged VMRs of ~17% - 36% and individual layer VMRs of ~10% - 60%.

The time period of 1.7 to 1 Ma to which the integration model is applied likely corresponds to a deeper section of NPLD layering than the upper ~500 m (assuming the NPLD accumulated at a nearly constant long-term rate of ~0.5 mm/yr (Laskar et al., 2002; Levrard et al., 2007; Hvidberg et al., 2012)). Nevertheless, by predicting periods of sustained polar water ice and dust accumulation with greater than average dust fraction, the results of this microphysics-inclusive investigation provide a closer match to observationally constrained VMRs than existing models.

Integration model results imply that the exchange of water between the polar reservoirs themselves may have exclusively driven NPLD growth / SPLD deflation during low-eccentricity epochs, which spanned ~one third of the NPLD's inferred formation history. The availability of SPLD water ice for sublimation prior to 1.7 million years ago is likely given the SPLD's estimated age of 10 – 100 myr (Plaut et al., 1988; Herkenhoff et al., 2000). Assuming a dust VMR of ~10% for the SPLD, its deflation by ~200 m over ~700 kyr is predicted to produce a ~20-m-thick dust lag deposit. However, thermal stability models by Bramson et al. (2017, 2019) predict that just a ~10 cm-thick dust lag may be sufficient to inhibit sublimation completely, implying that dust must be continuously removed, perhaps by dust lifting, to sustain long-term sublimation of the SPLD surface. In any case, the predicted accumulation of the NPLD at the expense of loss from the SPLD ~1.7 – 1 Ma is difficult to reconcile with the SPLD's current surface age estimate of ~10 million years. This work predicts that the sublimation of mid or high latitude water ice can provide an alternative source of NPLD water ice accumulation, but this results in NPLD accumulation rates as high as ~cm/yr, two orders of magnitude greater than the NPLD's inferred long-term accumulation rate of 0.1 mm/yr. The sublimation of these mid and high latitude sources (and corresponding polar accumulation rates) may be sufficiently reduced by the implementation of a dust lag scheme, which we have not explored in this investigation.

Even lower accumulation rates might be predicted if midlatitude sources are geographically confined to regions most conducive to midlatitude glaciation (Madeleine et al, 2009) or regions that exhibit the highest concentration of observed glacial features (Forget et al., 2017) rather than distributed over 5° latitude-wide planet-encircling bands that approximately encompass such regions. Furthermore, this work has not explored the possibility that alternative water ice sources, such as tropical glacial deposits or the release of water vapor from the buried latitude-dependent water ice mantle deposit, may have supplied NPLD buildup during low-eccentricity epochs.

Since they incorporate results from low obliquity polar cap source MGCM simulations, integration model simulations like that exhibited by Fig. 7 are predicted to accumulate $CO_2$ ice on the polar surfaces at obliquity minima. Integrated over low-obliquity excursions spanning tens of kyrs, predicted polar $CO_2$ ice accumulation produces a series of 50 m and 200 m-thick $CO_2$ deposits on the south and north polar surfaces, respectively. While the simulated production of such $CO_2$ deposits at low obliquities is consistent with Bierson et al. (2016) and with the presence of large $CO_2$ deposits at depth in Australe Mensae (Phillips et al., 2011), polar $CO_2$ accumulation is not nominally accounted for in our water ice and dust-focused work. The sublimation and possible removal of such deposits is inferred by their lack of detection within the NPLD and within the SPLD below ~1 km depth, but our model has no means of removing polar surface $CO_2$ once accumulated, preventing a complete assessment of their long-term evolution.

As discussed in section 2, the net annual polar deposition rates incorporated into the integration model are derived from MGCM simulations that all exhibit some degree of annual dust and water mass loss from their globally-integrated budgets. The maximum impact of this imprecise conservation on net annual polar surface deposition rates is assessed by assuming a "worst case" scenario in which the water and dust annually lost from the model is redeposited on one of the polar surfaces. The addition of leaked water to the polar surfaces tends to support the persistence of polar cap water ice reservoirs and thus further favor the selection of net annual polar accumulation from polar cap source MGCM models. Ingesting these "worst-case" rates into an integration model like that illustrated by Fig. 7 has the effect of thickening the ~225 m-thick deposit by ~33% (~75 m), increasing its bulk dust-to-water ice VMR from ~13% to ~27%, and increasing its minimum dust-to-water ice VMR from ~3% to ~7%. The dust-to-water ice VMR of dust-rich layers formed at obliquity maxima are roughly doubled while their thicknesses are increased by ~70%. The thicknesses of dust lag deposits formed at obliquity minima are increased by ~30%. Despite these effects, the deposits still exhibit a layer stratigraphy characterized by two dust-rich and two dust-poor layers per obliquity cycle formed via the same described layer formation mechanism.

## 5. Summary and Conclusions

We have utilized the NASA Ames Legacy Mars General Circulation Model and a parameterized integration model to investigate the sensitivity of Martian net annual polar deposition rates of water ice and dust to changes in planetary obliquity and surface water ice emplacement. The following conclusions can be drawn from this work:

1. MGCM results under zero orbital eccentricity conditions predict an annual net cap-to-cap exchange of water ice when both a north and south polar cap are prescribed (a plausible condition beginning in and possibly persisting after low-obliquity periods), with the rate of transfer greater with increased obliquity and the transfer direction dependent on obliquity (Table 1b). The accumulating polar region annually forms a surface deposit with a dust-to-water ice VMR of a few percent to 98%, while the losing polar region annually forms a dust lag deposit.

2. When midlatitude surface water ice deposits are prescribed (a plausible condition beginning in and possibly persisting after high-obliquity periods), MGCM results generally predict rapid net annual water ice accumulation at both poles at the expense of loss from these reservoirs. A combination of low atmospheric dust opacities (Fig. 6e-g) and high net annual polar water ice deposition rates (Table 1b) in these midlatitude source scenarios produces thicker but dust-depleted (< 1% VMR) annual polar deposits than simulations with polar sources.

3. MGCM results indicate that water deposition in the polar regions is primarily in the form of snowfall, while dust deposition is primarily in the form of nucleated "core" dust. This suggests that dust-water cycle coupling, specifically nucleation, is an important factor in determining PLD layer properties.

4. The integration model simulates northern deposits accumulated between ~1.7 - 1 MYA which exhibit an average accumulation rate close to the inferred rate of ~0.5 mm/yr for the NPLD's entire accumulation history. In the majority of simulations, both north and south polar water ice caps are present and persist over multiple obliquity cycles, with long-term accumulation of water ice in the north occurring at the expense of  long-term loss of water ice in the south. Persistent midlatitude sources would produce rates of accumulation in the north that would be too rapid and too sustained to account for the NPLD's current ~3 km thickness. However, this investigation has not considered the impact of dust lags or the emplacement of midlatitude ice in specific geographical locations (rather than planet encircling bands), which may reduce the rate of midlatitude ice-supplied NPLD buildup.

5. Integration models which produce average northern accumulation rates between 0.1 - 1 mm/yr produce a stack of multiple layers possessing bulk dust VMRs of ~5% to ~25% and individual layer VMRs of ~20% - 30%. While these column-averaged and individual layer VMRs are greater than the inferred bulk dust concentration of the entire NPLD (~5%), analytic models of the properties of bright reflectors in the upper 300 - 500 m of the NPLD (Lalich et al., 2019) similarly predict high column-averaged VMRs (~17% - 36%) and individual layer VMRs (~10% - 60%).

6. Integration models which produce average northern accumulation rates in the range 0.1 - 1 mm/yr typically exhibit a northern stratigraphy (Fig. 7d) characterized by two dust-rich layers per obliquity cycle: a 10's of meters-thick layer containing ~20% - 30% dust by volume that forms at high obliquity when both water ice deposition and dust deposition are high, and a ~0.5 m-thick dust lag deposit (100% dust) that forms at low obliquity when net annual removal of water ice occurs. The water and dust deposition components of this layer formation mechanism have an opposite correlation to obliquity compared to that of Hvidberg et al. (2012). However, this work, like Hvidberg et al. (2012), demonstrates the capability of obliquity variations to produce a stratigraphy similar to that observed in the NPLD.

**Acknowledgements**

This work benefited extensively from discussions with Robert Haberle and Jeffery Hollingsworth at NASA Ames Research Center. The authors thank Patricio Becerra and an anonymous reviewer for extensive and constructive reviews which greatly improved this final manuscript.

**Funding**

This work has been supported by National Aeronautics and Space Administration (NASA) Earth and Space Science Fellowship Award # NNX16AP37H and NASA Cooperative Agreement #NNX15AL51H

**Data Availability**

*Link to supplementary Table S1 here*

Table S1

**Table S1. Sensitivity of annual net polar water and dust deposition to all tested MGCM source and obliquity conditions.** Values are reported in mm/Earth year. Values are included for the present-day simulation (labelled "NRC") and for the 55 paleoclimate simulations (prefixed "1." - "11.", corresponding to the water ice source configurations listed in section 4.2.)

| Surface Ice Sources | Obliquity [deg] | North Water [mm/yr] | North Dust [mm/yr] | North VMR [%] | South Water [mm/yr] | South Dust [mm/yr] | South VMR [%] |
|---|---|---|---|---|---|---|---|
| NRC | 25.19 | 0.0633174 | 0.0105121 | 14.2384 | 0.00000 | 0.00515625 | 100.000 |
| 1. Caps 77.5-87.5 N&S | 15 | 0.000127747 | 0.00827850 | 98.4803 | -0.000161501 | 0.00436265 | 100.000 |
| | 20 | -0.00211528 | 0.00943684 | 100.000 | 0.000811166 | 0.00494891 | 85.9174 |
| | 25 | -0.0346565 | 0.0102957 | 100.000 | 0.0230923 | 0.00633447 | 21.5262 |
| | 30 | 1.05873 | 0.0312941 | 2.87095 | -1.00704 | 0.0214377 | 100.000 |
| | 35 | -0.440317 | 0.251772 | 100.000 | 0.135616 | 0.270887 | 66.6384 |
| 2. North Cap (77.5-87.5 N) | 15 | -1.62308e-05 | 0.00802636 | 100.000 | 1.07504e-05 | 0.00433364 | 99.7525 |
| | 20 | -0.00143400 | 0.00943274 | 100.000 | 0.000731505 | 0.00474244 | 86.6366 |
| | 25 | -0.0215924 | 0.0104218 | 100.000 | 0.00322244 | 0.00635053 | 66.3382 |
| | 30 | -0.0243392 | 0.0180290 | 100.000 | 0.00000 | 0.0176175 | 100.000 |
| | 35 | -0.0250196 | 0.260906 | 100.000 | 0.00000 | 0.271229 | 100.000 |
| 3. South Cap (77.5-87.5 S) | 15 | 9.63897e-05 | 0.00820776 | 98.8393 | -0.000154127 | 0.00437431 | 100.000 |
| | 20 | 0.000208507 | 0.00912553 | 97.7662 | -0.000937884 | 0.00479071 | 100.000 |
| | 25 | 0.00000 | 0.00991544 | 100.000 | -0.00635315 | 0.00605952 | 100.000 |
| | 30 | 0.731051 | 0.0339020 | 4.43190 | -0.670512 | 0.0248435 | 100.000 |
| | 35 | 0.00000 | 0.227308 | 100.000 | -0.229151 | 0.260684 | 100.000 |
| 4. Caps N&S (77.5-87.5 N/S) + Mid N&S (37.5-42.5 N/S) | 15 | 6.05601 | 0.0280907 | 0.461707 | 4.86126 | 0.00786392 | 0.161506 |
| | 20 | 8.85040 | 0.0191338 | 0.215725 | 7.29539 | 0.00609646 | 0.0834961 |
| | 25 | 12.1810 | 0.0126295 | 0.103574 | 10.3179 | 0.00419636 | 0.0406540 |
| | 30 | 11.6660 | 0.0105209 | 0.0901027 | 11.8861 | 0.00271455 | 0.0228328 |

| | | | | | | |
|---|---|---|---|---|---|---|
| | 35 | 2.02138 | 0.112840 | 5.28720 | 10.7781 | 0.0141079 | 0.130723 |
| 5. Caps N&S (77.5-87.5 N/S) + High N&S (57.5-62.5 N/S) | 15 | 0.489459 | 0.0149916 | 2.97187 | 0.307673 | 0.00808709 | 2.56115 |
| | 20 | 1.44800 | 0.0145066 | 0.991902 | 0.762599 | 0.00920247 | 1.19234 |
| | 25 | 2.94283 | 0.0139318 | 0.471183 | 2.08586 | 0.00802654 | 0.383333 |
| | 30 | 3.60441 | 0.172808 | 4.57500 | 2.20155 | 0.112507 | 4.86191 |
| | 35 | 1.57888 | 0.296652 | 15.8169 | 1.01739 | 0.268438 | 20.8767 |
| 6. Mid N&S (37.5-42.5 N/S) | 15 | 6.17731 | 0.0276877 | 0.446216 | 4.81159 | 0.00773249 | 0.160448 |
| | 20 | 8.73497 | 0.0196318 | 0.224245 | 7.40239 | 0.00626205 | 0.0845235 |
| | 25 | 11.3865 | 0.0128586 | 0.112802 | 10.4635 | 0.00437345 | 0.0417797 |
| | 30 | 12.5620 | 0.0103575 | 0.0823825 | 13.1643 | 0.00273636 | 0.0207819 |
| | 35 | -1.91583e-24 | 0.0821891 | 100.000 | 13.9558 | 0.00994537 | 0.0712125 |
| 7. High N&S (57.5-62.5 N/S) | 15 | 0.489685 | 0.0149407 | 2.96076 | 0.298620 | 0.00802270 | 2.61630 |
| | 20 | 1.30231 | 0.0150476 | 1.14225 | 0.745718 | 0.00902011 | 1.19513 |
| | 25 | 2.56424 | 0.0140421 | 0.544631 | 2.09245 | 0.00812153 | 0.386635 |
| | 30 | 4.40830 | 0.0132541 | 0.299760 | 3.13441 | 0.0131672 | 0.418327 |
| | 35 | 0.149390 | 0.302155 | 66.9158 | 0.00000 | 0.261382 | 100.000 |
| 8. Mid N (37.5-42.5 N) | 15 | 9.30801 | 0.233876 | 2.45104 | 8.05706 | 0.118187 | 1.44567 |
| | 20 | 8.11727 | 0.0188398 | 0.231558 | 2.56979 | 0.00970820 | 0.376361 |
| | 25 | 10.4147 | 0.0140741 | 0.134954 | 4.61842 | 0.00687256 | 0.148586 |
| | 30 | 11.1096 | 0.0101874 | 0.0916151 | 7.24715 | 0.00513832 | 0.0708510 |
| | 35 | 0.785755 | 0.122809 | 13.5168 | 12.5334 | 0.0154804 | 0.123361 |
| 9. High N (57.5-62.5 N) | 15 | 0.385338 | 0.0130392 | 3.27309 | 0.164906 | 0.00754657 | 4.37602 |
| | 20 | 1.13609 | 0.0138564 | 1.20497 | 0.487414 | 0.00902626 | 1.81820 |
| | 25 | 2.27433 | 0.0138287 | 0.604356 | 0.873601 | 0.00877263 | 0.994209 |
| | 30 | 3.04866 | 0.118399 | 3.73847 | 1.07253 | 0.0620372 | 5.46792 |
| | 35 | 0.0257928 | 0.297325 | 92.0175 | 0.00000 | 0.270771 | 100.000 |
| 10. Mid S (37.5-42.5 S) | 15 | 3.05525 | 0.0320102 | 1.03685 | 4.91122 | 0.00807212 | 0.164091 |
| | 20 | 4.64805 | 0.0243578 | 0.521311 | 7.08560 | 0.00663326 | 0.0935284 |

|  | 25 | 6.78643 | 0.0162772 | 0.239276 | 9.71781 | 0.00545629 | 0.0561158 |
|  | 30 | 8.44075 | 0.0119190 | 0.141009 | 11.6681 | 0.00322468 | 0.0276292 |
|  | 35 | 0.00000 | 0.0688083 | 100.000 | 11.4534 | 0.0106243 | 0.0926757 |
| 11. High S (57.5-62.5 S) | 15 | 0.187465 | 0.0148481 | 7.33917 | 0.223468 | 0.00728932 | 3.15887 |
|  | 20 | 0.396949 | 0.0158448 | 3.83843 | 0.605446 | 0.00844636 | 1.37587 |
|  | 25 | 0.632837 | 0.0147042 | 2.27077 | 1.82790 | 0.00752927 | 0.410219 |
|  | 30 | 3.98109 | 0.0715187 | 1.76476 | 1.62629 | 0.0516899 | 3.08049 |
|  | 35 | 0.135616 | 0.302770 | 69.0646 | -0.000267538 | 0.258579 | 100.000 |